\documentclass[twocolumn,aps,prx,superscriptaddress]{revtex4-1}  

\usepackage{dcolumn}   
\usepackage{bm}        
\usepackage{amssymb}   
\usepackage{hyperref}
\usepackage{amsmath}
\usepackage{amssymb}
\usepackage{mathtools}
\usepackage{soul,xcolor}

\newcommand{\vect}[1]{\mathbf{#1}}
\newcommand{\citeasnoun}[1]{Ref.~\citenum{#1}}

\newcommand{\figref}[1]{Fig.~\ref{#1}}
\newcommand{\figsref}[1]{Figs.~\ref{#1}}
\newcommand{\Figref}[1]{Figure~\ref{#1}}
\newcommand{\secref}[1]{Sec.~\ref{#1}}

\PassOptionsToPackage{normalem}{ulem}
\usepackage{ulem}

\usepackage{tikz,xcolor,hyperref}
\definecolor{lime}{HTML}{A6CE39}
\DeclareRobustCommand{\orcidicon}{
    \hspace{-2mm}
	\begin{tikzpicture}
	\draw[lime, fill=lime] (0,0) 
	circle [radius=0.16] 
	node[white] {{\fontfamily{qag}\selectfont \tiny ID}};
	\draw[white, fill=white] (-0.0625,0.095) 
	circle [radius=0.007];
	\end{tikzpicture}
	\hspace{-2.5mm}
}
\foreach \x in {A, ..., Z}{\expandafter\xdef\csname orcid\x\endcsname{\noexpand\href{https://orcid.org/\csname orcidauthor\x\endcsname}
			{\noexpand\orcidicon}}
}

\begin{document}

\title{Analytical criteria for designing multiresonance filters in scattering systems, with application to microwave metasurfaces}

\author{Mohammed Benzaouia\orcidA} \email[Corresponding author: ]{medbenz@mit.edu} \affiliation{Department of Electrical Engineering and Computer Science, Massachusetts Institute of Technology, Cambridge, MA 02139, USA.}
\author{John D. Joannopoulos} \affiliation{Department of Physics, Massachusetts Institute of Technology, Cambridge, MA 02139, USA.}
\author{Steven G. Johnson\orcidB} \affiliation{Department of Mathematics, Massachusetts Institute of Technology, Cambridge, MA 02139, USA.}
\author{Aristeidis Karalis\orcidC} \email[Corresponding author: ]{aristos@mit.edu} \affiliation{Research Laboratory of Electronics, Massachusetts Institute of Technology, Cambridge, MA 02139, USA.}

\begin{abstract}
We present general analytical criteria for the design of lossless reciprocal two-port systems, which exhibit prescribed scattering spectra $S(\omega)$ satisfying $S_{22}(\omega)=e^{i\varphi}S_{11}(\omega)$, including symmetric ($S_{22}=S_{11}$) or ``antimetric'' ($S_{22}=-S_{11}$) responses, such as standard filters (Butterworth, Chebyshev, elliptic, etc.). We show that the non-normalized resonant (quasi-normal) modes (QNMs) of all such two-port systems couple to the input and output ports with specific unitary ratios, whose relative signs determine the position of the scattering zeros on the real frequency axis. This allows us to obtain design criteria assigning values to the poles, background response, and QNM-to-ports coupling coefficients. Filter devices can then be designed via a well-conditioned nonlinear optimization (or root-finding) problem using a numerical eigensolver. 
As an application, we design multiple microwave metasurfaces configured for polarization-preserving transmission, reflective polarization conversion, or diffractive ``perfect anomalous reflection'', to realize filters that precisely match standard bandpass or bandstop filters of various types, orders and bandwidths, with focus on the best-performing elliptic filters.
\end{abstract}

\maketitle

\section{INTRODUTCTION}

High-order (multiresonance) filters---especially standard filters (SFs) of Butterworth, Chebyshev, or elliptic spectral shape~\cite{dimopoulos2011analog}---have been designed for many types of wave physics (electromagnetic~\cite{little1997microring, fan1998channel, manolatou1999coupling, popovic2006multistage, xiao2007highly, liu2011synthesis, dai2011circuit, bogaerts2012silicon, wu1995frequency, vardaxoglou1997frequency, munk2005frequency}, mechanical~\cite{morgan2010surface, colombi2016seismic, rostami2016acoustic, d2017mechanical, colquitt2017seismic}, etc.) by a variety of techniques, including brute-force optimization of the scattering (e.g., transmission) spectrum~\cite{manara1999frequency, kern2005design, bossard2006design, aage2017topology, jensen2005topology, jiang2013tailoring}, circuit theory in the microwave regime~\cite{liao2020quasi, li2013synthesis, costa2014overview, sarabandi2007frequency, bayatpur2008single, bayatpur2009metamaterial, mesa2015simplified}, and coupled-mode theory (CMT)~\cite{haus1984waves, fan2002analysis, suh2004temporal} for cascaded optical resonators~\cite{little1997microring, fan1998channel, manolatou1999coupling, popovic2006multistage, xiao2007highly, liu2011synthesis, dai2011circuit}. Circuit theory and CMT provide attractive semianalytical frameworks for filter design, but are restricted to systems composed of spatially separable components (either discrete circuit elements or weakly coupled resonators, respectively), while brute-force spectrum optimization faces several numerical challenges~\cite{aage2017topology, jensen2005topology}. To design ultra-compact filters, involving strongly coupled elements and spatially overlapping resonances, a precise, systematic and computationally tractable methodology is missing. In this article, we develop such a filter-design approach by deriving a minimal set of explicit \emph{analytical} criteria on the system resonances, applicable to all symmetric and ``antimetric''~\cite{maloratsky2003passive} filters, including SFs. To derive these conditions, we use the unitary and symmetric quasinormal-mode theory (QNMT) of the scattering matrix $S$ from \citeasnoun{QNMT} to derive the required coupling coefficients of the resonances (QNMs) to the input and output ports in conjunction with the net background response, in order to achieve multiple configurations for the zeros of the $S$ coefficients (generalizing previous work~\cite{li2010coupled, hsu2014theoretical}) and thus realize any desired SF. We apply our procedure to computationally design microwave metasurfaces with several two-port configurations, realizing filters that precisely match SFs of various orders, bandwidths, and types---especially optimal elliptic filters, which were demonstrated only approximately in the past~\cite{luo2007design, li2013synthesis, jiang2013tailoring, lv2019wide, liao2020quasi}.

Large-scale optimization (including a variety of inverse-design and machine-learning algorithms) is a powerful approach to design complex structures by optimizing thousands of degrees of freedom~\cite{jensen2011topology,Molesky2018}. However, if a filter optimization problem is formulated directly in terms of constraints on the transmission spectrum, it can face severe numerical challenges~\cite{aage2017topology, jensen2005topology}: the highly oscillatory nature of the transmission spectrum can trap optimizers in poor local optima, and stringent constraints (e.g., on stop- and passband transmission) can lead to very ``stiff'' optimization problems with slow convergence. For example, these issues forced one such effort~\cite{aage2017topology} to restrict the designs to spatially distinct resonators, as in CMT. However, when analytical solutions to \emph{parts} of the problem exist, the numerical side of the optimization can be rendered simpler and more robust. In particular, for filter design with given transmission-spectrum constraints, signal-processing theory analytically defines many such ``optimal'' standard filters, characterized  by various rational transfer functions with specified poles and zeros~\cite{dimopoulos2011analog}, the latter necessary to achieve a steep transition between the ``pass'' and ``stop'' frequency bands. Therefore, when designing physical filters, it is advantageous to exploit these analytical solutions. An exact methodology, called network synthesis, was developed to implement these SFs in the extreme quasistatic (subwavelength) limit, where structures can be modeled precisely by networks of discrete elements, as in electronic circuits~\cite{dimopoulos2011analog}. In the other limit of structures spanning multiple wavelengths, the simple mapping between coupled resonators and transfer-function poles has made temporal CMT~\cite{haus1984waves, fan2002analysis, suh2004temporal} a popular design tool, especially for (high-order) optical add-drop filters~\cite{fan1998channel, little1997microring, popovic2006multistage, xiao2007highly, manolatou1999coupling, liu2011synthesis}, most of which are only Chebyshev or Butterworth filters with no transmission zeros, using a symmetric topology. However, in the intermediate limit of physical structures with size of the order of the wavelength or only a few times smaller (metamaterials), no complete filter-design methodology exists. Equivalent circuits with elements calculated from analytical expressions are not accurate and usually serve only as initial guess for trial-and-error design~\cite{li2013synthesis, sarabandi2007frequency, bayatpur2008single, liao2020quasi, costa2014overview}. For better accuracy, the effective element values should be obtained by fitting to the actual spectral response~\cite{bayatpur2008single, liao2020quasi, costa2014overview}, which is not practical for optimization (especially for sharp spectra). Moreover, these circuits often become overly complicated~\cite{mesa2015simplified, liao2020quasi}, they change for each different structural topology~\cite{bayatpur2009metamaterial} or, worse, they fail to provide any adequate model (as is typically the case in dielectric photonic structures). Therefore, network synthesis may be useful for the intuitive choice of an appropriate system topology but not for the calculation of its exact parameters. CMT, on the other hand, is typically based on weakly coupled resonators and the knowledge of the ``uncoupled'' modes of the system~\cite{suh2004temporal}, but neither of these conditions usually hold for wavelength-scale structures with multiple strongly intercoupled or overlapping resonances~\cite{popovic2006coupling}. Still missing has been a unified, physics-independent, set of exact conditions for the precise design of filters with multiple zeros that can be fed as a smooth objective to optimization algorithms. Using our QNMT of~\citeasnoun{QNMT} (whose main results are summarized in \secref{sec-2port-general}), in \secref{sec-filter-design} we derive such simple and general rules to design SFs using eigenmode solvers. In particular, we show that the resonant QNM fields of \emph{all} lossless reciprocal two-port systems with symmetric ($S_{22}=S_{11}$) or ``antimetric'' ($S_{22}=-S_{11}$)~\cite{maloratsky2003passive} response couple to the input and output ports with specific unitary ratios, whose relative signs determine the position of the scattering zeros. Thus, for filter design, apart from the obvious matching of system resonant frequencies to the desired filter's complex poles, we explain that, to also obtain the desired-filter zeros, these ratios must be enforced for the critical filter resonances and the remaining QNMs must add up to a required overall background response.

As an application of our theory, we design microwave frequency selective surfaces (FSSs), which are usually used to implement spatial (wave) filters for communication antennas, radars, radomes ~\cite{wu1995frequency, vardaxoglou1997frequency, munk2005frequency}, lenses~\cite{al2011wideband,li2012broadband} etc. FSSs typically take the form of two-dimensional periodic metal-dielectric arrays exhibiting specific frequency-dependent transmission or reflection under plane-wave excitation. While older designs were based on wavelength-sized unit cells (as in typical antenna design), the use of subwavelength dimensions to form metasurfaces has attracted much attention in the past decade due to multiple advantages, such as higher unit-cell density and smaller angular sensitivity~\cite{sarabandi2007frequency, bayatpur2008single, glybovski2016metasurfaces}. An important design challenge in frequency-selective metasurfaces is the ability to obtain specific high-order frequency responses using their strongly intercoupled subwavelength resonances, attempted usually through multilayer FSSs. Most previous efforts have been based on effective-circuit models~\cite{liao2020quasi, li2013synthesis, costa2014overview, sarabandi2007frequency, bayatpur2008single, bayatpur2009metamaterial}. The basic FSS building blocks are metallic patches with gaps (effective capacitors $C$) and apertures/loops (effective inductors $L$) that can be combined to make effective $LC$ resonators. Then, the shape, size and arrangement of patches and apertures in the FSS dielectric and metallic sheets are designed to accomplish the necessary circuit topology and element values for the transmission desired. While such circuit models can give a good physical intuition about the expected response of a FSS, they are too approximate and less flexible for a precise design method (as explained above). This is why, although particular attention has been given to the design of elliptic filters, most previous efforts have only achieved an approximate ``\emph{quasi}elliptic'' response~\cite{luo2007design, jiang2013tailoring, lv2019wide, li2013synthesis, liao2020quasi}. In \secref{sec_fss}, we first discuss the relation between QNMT and effective-circuit models to motivate appropriate structural-topology choices for different filters and scattering-zero placements. Then, following our systematic filter-design procedure, we implement polarization-preserving transmissive microwave metasurfaces that exhibit, for a normally incident plane wave, transmission spectra matching SF responses of various orders, bandwidths and types. Notably, we demonstrate second- and third-order elliptic filters for both bandpass and bandstop behaviors. We show that, in some cases, even though symmetric performance is desired, structural asymmetry should be used, while conversely, in cases where the ideal performance is antimetric, we also present approximate symmetric solutions. Lastly, to highlight the generality of our method, we also design metasurface SFs for different two-port configurations: a reflective polarization converter~\cite{Grady2013, Feng2013, Zhang2016, Loncar2018, Karamirad2021} and a diffractive ``perfect anomalous reflector"~\cite{Hessel1975, Tretyakov2017, Radi-Alu2017, Eleftheriades2018, Epstein2019}. These have been previously demonstrated mostly at single frequencies, while here we match a desired spectral response. The designed metasurfaces are compatible with fabrication by printed-circuit board technology, and also offer potential electrical tunability.

Details regarding the optimization setup, including objectives and algorithm, are provided in \secref{sec:design_optimization}. Therein, we also demonstrate the superiority of our QNMT-based design method compared to the common approach of brute-forcing the desired spectrum at few key frequencies: for three different initial structures, our optimization method always converged to structures matching the target spectral response, while the direct frequency-domain optimization always failed to find a good solution.

\section{$S$-MATRIX OF LOSSLESS RECIPROCAL TWO-PORT SYSTEMS}
\label{sec-2port-general}

\begin{figure}
\includegraphics[width=1\columnwidth,keepaspectratio]{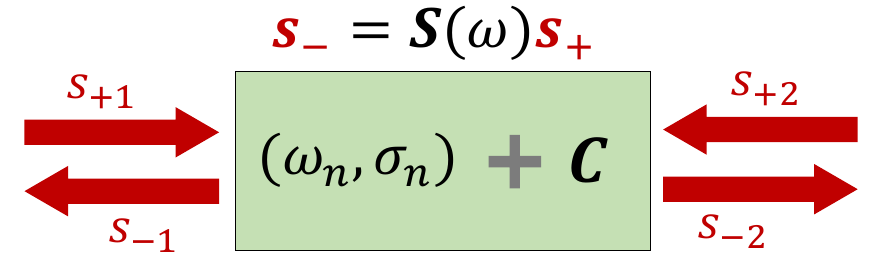}
\caption{A lossless reciprocal two-port scattering system excited at frequency $\omega$, with input and output amplitudes respectively $s_{\pm p}$, related by the $S$-matrix through $s_-=Ss_+$. The system supports high-$Q$ quasi-normal modes (QNMs) with frequencies $\omega_n$ and port-coupling ratios $\sigma_n$, while low-$Q$ resonances create an effective background response $C$.}
\label{Fig-2port} 
\end{figure}

We consider a linear time-independent reciprocal system, without material absorption or gain (although these could easily be included perturbatively~\cite{QNMT}), coupled to radiation only via two ports. These are used as channels for an incoming excitation at frequency $\omega$ and outgoing scattered waves, described by a $2\times2$ scattering matrix $S$ (\figref{Fig-2port}). Here, we summarize some key properties of $S$ and its QNMT model, derived in~\citeasnoun{QNMT}, that we need 
in later sections.

To begin with, for port modes whose transverse field does not depend on frequency (such as plane waves or dual-conductor TEM microwave modes), the scattering matrix can typically be written as $S=e^{i\tau\omega}S'e^{i\tau\omega}$, where $S'$ is a ``proper'' rational function corresponding to ports' reference cross sections taken on the surface of the scatterer and $\tau$ is a constant diagonal matrix with real positive elements corresponding to the propagation delay through the ports (see Appendix~A and section IV-C of \citeasnoun{QNMT}). Hereafter, we always consider those unique reference cross sections and drop the prime notation so that $S$ is rational and any propagation phase can be easily added in the end.

Moreover, recall (Ref.~[\citealp{QNMT}, Appendix~E]) that (i) the poles of $S$ appear in pairs ($\omega_n,-\omega_n^*$) due to realness [$S^{*}(i\omega)=S(-i\omega^{*})$]; (ii) the zeros of $S_{21}=S_{12}$ can only appear as complex quadruplets ($\omega_{o},\omega_{o}^{*},-\omega_{o},-\omega_{o}^{*}$), real or imaginary pairs ($\omega_{o},-\omega_{o}$), or at $\omega_{o}=0$; and (iii), for each zero pair ($\omega_{o},-\omega_{o}^{*}$) of $S_{11}$, ($-\omega_{o},\omega_{o}^{*}$) is a zero pair of $S_{22}$. These restrictions imply that $S_{pq}$ is a rational function of $i\omega$ with real coefficients and, in particular, that the numerator of $S_{21}$ is a polynomial of $\omega^{2}$ with real coefficients, optionally with multiplicative $i\omega$ factors.

The system poles correspond to resonant QNMs, which can be obtained using a numerical eigensolver. The modes with high ``quality factor'' $Q$ have frequencies $\omega_n$ and coupling coefficients to the ports $p=1,2$ equal to $D_{pn}$, which can be computed as an overlap surface integral between the $n$-QNM field and the $p$-port mode at the boundary of the scatterer, as explained in detail in \citeasnoun{QNMT}. Their ratios $\sigma_n = D_{2n}/D_{1n}$ do not depend on the normalization of the QNMs (Ref.~[\citealp{QNMT}, Appendix~D]). Moreover, any system low-$Q$ resonances $\{\omega_n^C,\sigma_n^C\}$ can admit a simplified description in terms of an effective ``background'' response between the two ports, quantified by a background scattering matrix $C$~\cite{QNMT}. When these background QNMs have $Q\rightarrow 0$, they give a frequency-independent unitary symmetric $C$. In this case, our formulation from \citeasnoun{QNMT} shows that enforcing energy conservation (unitary $S$) gives
\begin{equation}
\begin{gathered}
S(\omega)=\bar{S}(\omega)C\\ \bar{S}_{\{\omega_{n},\sigma_n\}} (\omega) =I+\sum_{n=1}^{N}\frac{\bar{S}^{(n)}}{i\omega-i\omega_{n}}\\
\bar{S}^{(n)}_{pq}=\sigma_{pn}\sum_{l=1}^{N}M^{-1}_{\;nl}\sigma_{ql}^{*},\\  M_{nl}=\frac{1+\sigma_{l}\sigma_{n}^{*}}{i\omega_{l}-i\omega_{n}^{*}},\;\;
\left(\begin{array}{l}\sigma_{1n}=1\\\sigma_{2n}=\sigma_{n}\end{array}\right),
\end{gathered}\label{S_TC}
\end{equation}
where $\sigma_n$ are further fine-tuned from the computed values using a simple constrained optimization, in order to also satisfy the reciprocity condition (symmetric $S$)
\begin{equation}
[\bar{S}^{(n)}C]_{21}=[\bar{S}^{(n)}C]_{12}.
\label{eq:TC_reciprocity}
\end{equation}
The matrix $C$ itself can be computed as $C=-\bar{S}_{\{\omega_n^C,\sigma_n^C\}}$. In practice, the background $Q$s are small but nonzero, so $C(\omega)$ is slowly varying but not constant; nevertheless, Eq.~(\ref{S_TC}) provides a good approximation for $S$. The distinction between high- and low-$Q$ modes is, in fact, somewhat arbitrary and based mainly on computational convenience. In the limit where one includes all modes in $\bar{S}$, then $C=-I$.

It can be easily shown \cite{QNMT} that Eqs.~(\ref{S_TC}) define a one-to-one mapping $\{\omega_n,\sigma_n\} \leftrightarrow \bar{S}(\omega)$ and that, for a constant $\gamma=\pm1$ (extended to any complex $\gamma=e^{i\varphi}$ in \secref{sec_approx_solutions}),
\begin{equation}
\begin{array}{c}
\bar{S}_{11\{\omega_{n},\sigma_{n}/\gamma\}}=\bar{S}_{11\{\omega_{n},\sigma_{n}\}},\;
\bar{S}_{12\{\omega_{n},\sigma_{n}/\gamma\}}=\gamma \bar{S}_{12\{\omega_{n},\sigma_{n}\}} \\
\bar{S}_{21\{\omega_{n},\sigma_{n}/\gamma\}}=\bar{S}_{21\{\omega_{n},\sigma_{n}\}}/\gamma,\;
\bar{S}_{22\{\omega_{n},\sigma_{n}/\gamma}\}=\bar{S}_{22\{\omega_{n},\sigma_{n}\}}.
\end{array}\label{eq:T-phase-dependence}
\end{equation}
and, by swapping ports $1\leftrightarrow2$,
\begin{equation}
\bar{S}_{11\{\omega_{n},\sigma_{n}\}} = \bar{S}_{22\{\omega_{n},1/\sigma_{n}\}}, \; \bar{S}_{21\{\omega_{n},\sigma_{n}\}} = \bar{S}_{12\{\omega_{n},1/\sigma_{n}\}}.
\label{eq:T-12-21}
\end{equation}

\section{QNMT-DERIVED ANALYTICAL CRITERIA FOR FILTER DESIGN}
\label{sec-filter-design}

Our goal is to design physical two-port systems with multiresonance network-synthesis filter responses, specifically, $N$th-order bandpass and bandstop filters of a finite bandwidth, which are given as rational functions of frequency $H(\omega)$ with specified $2N$ complex poles [appearing as $N$ pairs ($\omega_n,-\omega_n^*$)], $2N$ zeros (abiding by the restrictions of the previous section) and an overall constant. In the case of the standard-filter (SF) approximations, these are given through ``textbook'' analytical expressions~\cite{dimopoulos2011analog}.

It is obvious that the complex resonant frequencies $\omega_n$ of the physical system must match the complex poles of the desired filter. In this work, we show how to also enforce the desired zeros in the system response, by deriving the corresponding $\sigma_n$ coefficients and the matrix $C$. Specifically, $|C_{21}|$ must match the desired filter background transmission and then, for exact SFs, we find that the ratios of couplings of the QNMs' fields to the two ports must be $\sigma_n=\pm 1$ for $N$ odd or $\sigma_n=\pm i$ for $N$ even, with alternating signs for consecutive modes. We also explain that good approximate solutions can be obtained, if an overall common phase for all $\sigma_n$ is allowed, according to Eqs.~(\ref{eq:condition-a}--\ref{eq:condition-d}) below.

\subsection{Symmetric and antimetric filters}

We explained that, for a general lossless reciprocal two-port system, the zeros of $S_{11}$ and $S_{22}$ are conjugates of each other, but they do not necessarily coincide. In this article, we are interested in the special cases of filters where they do coincide, so that these zeros can only appear as complex quadruplets ($\omega_{o},\omega_{o}^{*},-\omega_{o},-\omega_{o}^{*}$), real or imaginary pairs ($\omega_{o},-\omega_{o}$), or at $\omega_{o}=0$. Their numerator is then also (as is always true for $S_{21}$) a polynomial of $\omega^{2}$ with real coefficients, optionally with multiplicative $i\omega$ factors. These cases include, in particular, common practical amplitude filters, for which all zeros of reflection ($S_{11}$ and $S_{22}$), corresponding to full transmission, lie on the real frequency axis or at infinity. To satisfy realness [$S^{*}(i\omega)=S(-i\omega^{*})$], this class of filters is collectively described by the condition $S_{22}=\pm S_{11}$, namely they are either symmetric or ``antimetric''~\cite{maloratsky2003passive} (note~\footnote{This should not be confused with a symmetric $S$ matrix, which holds for a \emph{reciprocal} system.}). Energy conservation and reciprocity then force $\sqrt{\gamma} S_{11}S^*_{21}$ to be purely imaginary for $\gamma \equiv S_{22}/S_{11} = \pm1$, corresponding to an odd (+) odd or even (-) number of $i\omega$ factors in the numerator of $S_{11}$ or $S_{21}$.

The most important subclass comprises the SF approximations of the ideal rectangular filter~\cite{dimopoulos2011analog}: Butterworth (flat passband and stopband), Chebyshev (equiripple passband, flat stopband), inverse Chebyshev (equiripple stopband, flat passband), and elliptic (equiripple passband and stopband) (see Fig.~\ref{Fig-grid3} in \secref{subsub:third-order-bandpass}). For an $N\mathrm{th}$-order Butterworth or Chebyshev transmission bandpass filter, $S_{21}$ has $N$ zeros at $\omega=0$ ($N$ $i\omega$ factors in numerator) and $N$ zeros at $\omega\rightarrow\infty$ ($2N$ zeros total). For $N^\mathrm{th}$-order inverse Chebyshev or elliptic filters, which have zeros at finite real frequencies, $S_{21}$ still has one zero at $\omega=0$ and one at $\omega\rightarrow\infty$ for $N$ odd, while all $2N$ zeros are finite for $N$ even (no $i\omega$ factors). For a transmission bandstop filter, the same observations hold instead for $S_{11}$. In all SF cases, we conclude that $S_{11}S^*_{21}$ is purely imaginary ($\gamma=1$) if $N$ is odd, and purely real ($\gamma=-1$) if $N$ is even.

\subsection{Conditions on $C$ and $\sigma_n$}
\label{subsec:conditions}

A partial-fraction expansion of the desired network-synthesis symmetric or antimetric filter expresses $H(\omega)$ in terms of the complex poles, their residues, and a direct term $t$ (which gives the limiting response at high frequencies according to the filter's type). For an actual physical system, the $S=\bar{S}C$ formulation of Eq.~(\ref{S_TC}) assumes that $C$ is approximately constant over the finite bandwidth of interest, where $C$ can generally be complex. Far from the high-$Q$ resonances ($\omega\gg\left|\omega_{n}\right|$), Eq.~(\ref{S_TC}) then dictates that $\bar{S}\rightarrow I$ and thus $S\rightarrow C$. Therefore, for a transmission filter, one must design $|C_{21}|=t$, and to also ensure that $|C_{21}|$ is indeed fairly constant within the filter operational bandwidth, it may often be useful to impose additional constraints (for example, $d^k|C_{21}|/d\omega^k\approx0$ for $k=1,2,...$ at the filter center frequency $\omega_\mathrm{c}$). The formula $C=-\bar{S}_{\{\omega_n^C,\sigma_n^C\}}$, which is used to calculate $C$, can also provide design intuition for the topology of the structure, where appropriate low-$Q$ modes are utilized to get the desired $C$, as we see in practical examples later. [Note that, during structural optimization, it may be difficult to find all the low-$Q$ modes contributing to $C$, when the relevant region of the complex plane is polluted by other noncontributing modes, such as diffraction branch cuts, perfectly matched layer (PML) modes, etc. (see the examples later and Ref.~[\citealp{QNMT}, Appendix~F]). Thankfully, $C$ can always also be calculated as $C=\bar{S}^{-1}S$, where $\bar{S}$ from Eq.~(\ref{S_TC}) includes only the filter-relevant high-$Q$ modes and $S$ is obtained via (additional) direct simulation of the structure (with the ports referenced at the scatterer boundary), but some numerical precautions should be taken (see section~\ref{sec:design_optimization}~A).]

For the class of filters of interest with $S_{22}=\gamma S_{11}$ ($\gamma=\pm1$), $S(\omega\gg|\omega_{n}|)\rightarrow C$ means that the unitary symmetric $C$ also satisfies $C_{22}=\gamma C_{11}$ and that $\sqrt{\gamma}C_{11}C^*_{21}$ is purely imaginary. Now, since $\bar{S}=SC^{-1}$, we can write: 
\begin{equation}
\bar{S}=\frac{1}{|C|}\begin{pmatrix} S_{11}\gamma C_{11}-S_{21}C_{21} & -S_{11}C_{21}+S_{21}C_{11}\\
 S_{21}\gamma C_{11}-\gamma S_{11}C_{21} & -S_{21}C_{21}+\gamma S_{11}C_{11}
\end{pmatrix}
\end{equation}
so that $\bar{S}_{11}=\bar{S}_{22}$ and $\bar{S}_{21}=\gamma \bar{S}_{12}$. Furthermore, for the QNM parameters $\{\omega_n,\sigma_n\}$ of this $\bar{S}$, using Eqs.~(\ref{eq:T-phase-dependence}, \ref{eq:T-12-21}) for the dependence of $\bar{S}$ on $\sigma_{n}$,
\begin{equation}\label{eq:T_symmetry}
\begin{array}{c}
\bar{S}_{11\{\omega_{n},1/\sigma_{n}\}}=\bar{S}_{22\{\omega_{n},\sigma_{n}\}}=\bar{S}_{11\{\omega_{n},\sigma_{n}\}}=\bar{S}_{11\{\omega_{n},\sigma_{n}/\gamma\}}\\
\bar{S}_{21\{\omega_{n},1/\sigma_{n}\}}=\bar{S}_{12\{\omega_{n},\sigma_{n}\}}=  \bar{S}_{21\{\omega_{n},\sigma_{n}\}}/\gamma=\bar{S}_{21\{\omega_{n},\sigma_{n}/\gamma\}}.
\end{array}
\end{equation}
The same result applies to $\bar{S}_{22}$ and $\bar{S}_{12}$, so we obtain $\bar{S}_{\{\omega_{n},1/\sigma_{n}\}} = \bar{S}_{\{\omega_{n},\sigma_{n}/\gamma\}}$. From the one-to-one mapping mentioned earlier, we conclude that $1/\sigma_{n}=\sigma_{n}/\gamma$. Therefore, for a lossless reciprocal two-port system, 
\begin{equation}
S_{22}=\gamma S_{11} \Leftrightarrow \sigma_{n}=\pm\sqrt{\gamma},\label{r-equation}
\end{equation}
so that all modes have $\sigma_{n}=\pm 1$ for a symmetric filter ($\gamma=1$), while $\sigma_{n}=\pm i$ for an antimetric filter ($\gamma=-1$). (This generalizes the well-known CMT result for a single resonance, where transmission reaches 1 only for equal decay rates into the two ports~\cite{JoannopoulosJo08-book}.) Moreover, with this $\sigma_{n}$ choice, $i\sigma_{n}C_{11}C^*_{21}$ is purely real.

When $C$ is exactly constant over all frequencies (as it is for exact SFs), it must be real, to satisfy the realness condition. Consistently with $i\sigma_{n}C_{11}C^*_{21}$ real, odd-order SFs have $\sigma_{n}=\pm 1$ and $C_{11}C_{21}=0$, while even-order SFs have $\sigma_{n}=\pm i$.

As detailed in Appendix~\ref{app_transfer_matrix}, Eq.~(\ref{r-equation}) can be also derived using general arguments based on the transfer matrix. However, the scattering-matrix QNMT we used here further helps specify the choice of $\sigma$ signs to enforce the desired positions of $S$-coefficients' zeros, as we show in the remainder of this section.

\subsection{Types of filters}
\label{sec_filter_types}

For each mode $n$, the appropriate choice of sign for $\sigma_n$ in Eq.~(\ref{r-equation}) depends on the specific filter type that is being designed. We find the adequate choice analytically in the limit of large $Q$s, or more specifically when $\Gamma_{n,l}\ll|\Omega_{n}-\Omega_{l}|$ for $\omega_{n}=\Omega_{n}-i\Gamma_{n}$. Under a such condition, the matrix $M$ is dominated by its diagonal terms $M_{nn}=(1+\left|\sigma_{n}\right|^{2})/2\Gamma_{n}$, so Eq.~(\ref{S_TC}) becomes 
\begin{equation}
\bar{S}_{pq}\approx\delta_{pq}+\sum_{n}\frac{\Gamma_{n}}{i\left(\omega-\Omega_{n}\right)-\Gamma_{n}}\frac{2\sigma_{pn}\sigma_{qn}^{*}}{1+\left|\sigma_{n}\right|^{2}},
\end{equation}
with $\sigma_{1n}=1$, $\sigma_{2n}=\sigma_{n}$. Then, away from the resonances $(\Gamma_{n}\ll|\omega-\Omega_{n}|)$ and to lowest order in $\Gamma_n$, further using $\left|\sigma_{n}\right|=1$ from Eq.~(\ref{r-equation}), transmission is 
\begin{equation}
S_{21}\approx C_{21}-i\sum_{n}\frac{\Gamma_{n}}{\omega-\Omega_{n}}\left(\sigma_{n}C_{11}+C_{21}\right).\label{eq:S21-for-r1}
\end{equation}
As the overall background transmission $C_{21}\approx S_{21}\left(\omega\gg\Omega_{n}\right)$ determines the filter type, we will study its different cases separately.

\begin{figure}
\includegraphics[width=1\columnwidth,keepaspectratio]{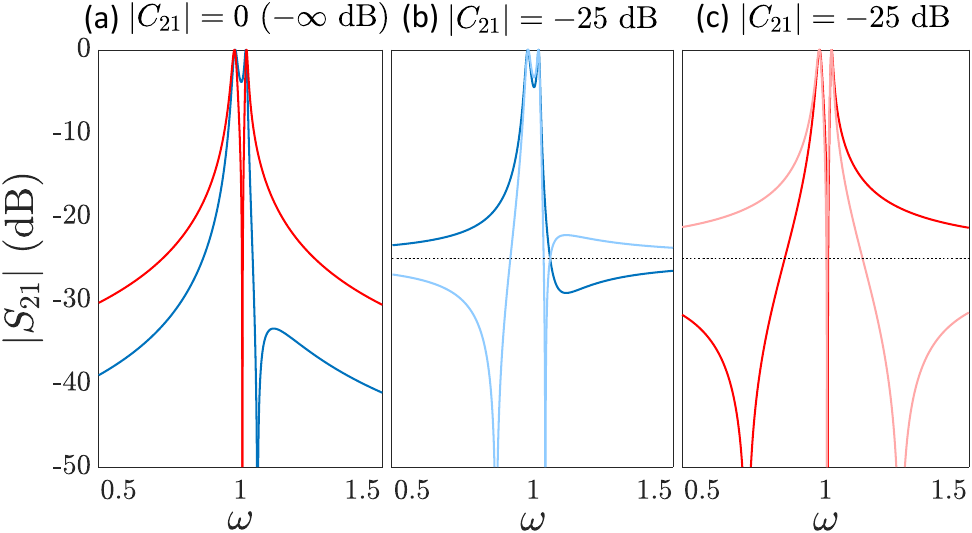} 
\caption{Second-order filter responses, using QNM expansions of the form $S=\bar{S}C$, with two modes of frequencies $\omega_n=(0.98-0.01i,1.02-0.005i)$ and coupling ratios $(\pm\sigma,\pm\sigma)$ in $\bar{S}$, and with unitary reciprocal $C=\begin{pmatrix}i\sigma^*r & t\protect\\t & i\sigma r \end{pmatrix}$ where $r=\sqrt{1-t^{2}}$ and such that $i\sigma\beta=-1$ and $C_{22}=\sigma^2 C_{11}\;(\Leftrightarrow S_{22}=\sigma^2 S_{11}$), for different values of real background transmission $t$ indicated in the plots. We use the RWA of ignoring negative-frequency poles, so the amplitudes of $S$-matrix coefficients are exactly the same for any complex $\sigma=\sqrt{\gamma}=e^{i\varphi/2}$. A zero always occurs between modes of same coupling ratio (red), so filters with no zeros between the poles require opposite signs of the ratios (blue). Including negative-frequency poles would result to the same qualitative behavior and only slightly change the response away from the resonances.}
\label{Fig-C-background} 
\end{figure}

\subparagraph*{Case (a) $C_{21}=0\Leftrightarrow|C_{11}|=1$:} This is a bandpass filter with zero transmission at $\omega\rightarrow\infty$. From Eq.~(\ref{eq:S21-for-r1}), we have $S_{21}\propto\sum_{n}\Gamma_{n}\sigma_{n}/(\omega-\Omega_{n})$, which, under condition Eq.~(\ref{r-equation}), is proportional to a real function that can be easily used to determine the placement of its zeros. As an example, we look at the simple scenario of two modes with $\Omega_{1}<\Omega_{2}$ and calculate the zero at
\begin{equation}
\omega_{o}\approx\frac{\Omega_{2}\Gamma_{1}\sigma_{1}+\Omega_{1}\Gamma_{2}\sigma_{2}}{\Gamma_{1}\sigma_{1}+\Gamma_{2}\sigma_{2}}.\label{eq:zero_for2modes}
\end{equation}
One can easily confirm that, when $\sigma_{1}=\sigma_{2}$, the zero appears between the modes ($\Omega_{1}<\omega_{o}<\Omega_{2}$), a feature often observed in interference phenomena, such as electromagnetically induced transparency (EIT)~\cite{maleki2004tunable, li2010coupled}. When $\sigma_{1}=-\sigma_{2}$, it appears on the side of the mode with the smallest loss rate ($\omega_{o}<\Omega_{1}$ if $\Gamma_{1}<\Gamma_{2}$ and $\omega_{o}>\Omega_{2}$ if $\Gamma_{2}<\Gamma_{1}$), while there is no zero if $\Gamma_{2}=\Gamma_{1}$ (explaining the lack of transmission zeros predicted by traditional CMT for two equal-loss coupled resonances~\cite{suh2004temporal} and in symmetric ``Fabry-Perot'' systems where all $\Gamma$s are the same~\cite{haus1984waves}). These points are illustrated in \Figref{Fig-C-background}(a).

These conclusions can be extended to the scenario of multiple high-$Q$ modes: a real zero always occurs between two consecutive modes with the same $\sigma_{n}$, there can only be an even (or zero) number of real zeros between two consecutive modes with opposite $\sigma_{n}$, and, below the lowest mode or above the highest mode, a zero can exist only if there is at least one change in the $\sigma$ signs. Examples of such high-order filters are given in Appendix~\ref{app_nonstandard_spectra}.

For the SFs with no transmission at infinity, such as a Butterworth, Chebyshev, odd-order inverse Chebyshev, or odd-order elliptic, where the transmission zeros are always outside the passband, it is necessary to design $\sigma_{n}$ with alternating signs, namely 
\begin{equation}\label{eq:condition-a}
\sigma_{n}=\pm\sqrt{\gamma}\left(1,-1,1,-1,...\right)=\pm\sqrt{\gamma}\left(-1\right)^{n-1}.   
\end{equation}

\subparagraph*{Case (b) $0<|C_{21}|\ll|C_{11}|<1$:} This is a bandpass filter with finite small transmission at $\omega\rightarrow\infty$. From Eq.~(\ref{eq:S21-for-r1}), we have $S_{21}\approx C_{21}-iC_{11}\sum_{n}\Gamma_{n}\sigma_{n}/(\omega-\Omega_{n})$. If we let $\beta=C_{11}C^*_{21}/|C_{11}C_{21}|$ then, from the previous discussion, $i\sigma_{n}\beta=\pm i\sqrt{\gamma}\beta=\pm 1$. Therefore, $S_{21}$ is proportional to a real expression, whose zeros can be easily predicted, as in the previous case. In particular, a real zero always occurs between two consecutive modes of the same $\sigma_{n}$. Moreover, there is an odd number of real zeros below the lowest mode, when $i\sigma_{1}\beta=-1$, and an even (or zero) number when $i\sigma_{1}\beta=1$, with a similar result  above the highest mode but for opposite signs of $i\sigma_{N}\beta$. This is simply illustrated in \figsref{Fig-C-background}(b,c) for two high-$Q$ modes and in Appendix~\ref{app_nonstandard_spectra} for higher filter orders.

For SFs, in this case inverse Chebyshev or elliptic of even order $N$ ($\gamma = -1\Leftrightarrow\beta=\pm1$), with a total of $N/2$ positive-frequency transmission zeros on each side of the passband, it is necessary to have 
\begin{equation}\label{eq:condition-b}
\sigma_{n} = \frac 1 \beta \left(-1\right)^{n-1}i^{N-1}, \; \beta=\frac{C_{11}C^*_{21}}{|C_{11}C_{21}|}.  
\end{equation}

\subparagraph*{Case (c) $0<|C_{11}|\ll|C_{21}|<1$:} This is a bandstop filter with finite small reflection at $\omega\rightarrow\infty$. Similar analysis by considering $S_{11}$ dictates, for even-order inverse Chebyshev or elliptic SFs,
\begin{equation}\label{eq:condition-c} \sigma_{n}=\frac 1 \beta \left(-1\right)^{n-1}i^{N+1}, \; \beta=\frac{C_{11}C^*_{21}}{|C_{11}C_{21}|}. \end{equation} 

\subparagraph*{Case (d) $C_{11}=0\Leftrightarrow|C_{21}|=1$:} This is a bandstop filter with zero reflection at $\omega \rightarrow \infty$ and a SF implementation requires again simply 
\begin{equation}\label{eq:condition-d}
\sigma_{n} = \pm\sqrt{\gamma}\left(-1\right)^{n-1}.
\end{equation}

By designing $\sigma_{n}$ to satisfy the appropriate condition from Eqs.~(\ref{eq:condition-a}--\ref{eq:condition-d}) according to the filter type, the pole residues in the partial-fraction expansion of $H(\omega)$ are also matched and thus the filter design is complete.

\subsection{$S_{22}(\omega)=e^{i\varphi}S_{11}(\omega)$ filters}
\label{sec_approx_solutions}

To exactly satisfy realness, we recall that negative-frequency modes are necessary and only $\gamma=\pm1$ is allowed. However, for some systems, realness may not be a strict condition. For example, for filters with high-$Q$ modes, the response can be well approximated (at positive frequencies $\omega$) by the well-known rotating-wave approximation (RWA) of including only positive-frequency modes in QNMT. In this case, all previous results hold for any complex phase factor $\gamma=e^{i\varphi}$. Therefore, in filter design, Eqs.~(\ref{eq:condition-a}--\ref{eq:condition-d}) permit $\sigma_{n}$ to be tuned to the desired values up to an overall common phase factor (expressed via $\gamma$ or $\beta=\pm i/\sqrt{\gamma}$). Then, the resulting filters, even after also including negative modes with their corresponding $\sigma^*_{n}$ to satisfy realness, will be good approximations of SFs within the bandwidth of interest.

As seen from Eq.~(\ref{eq:T-12-21}), for $\sigma=\pm1$ ($\gamma=1$), $\bar{S}$ is always a symmetric matrix, so $S=\bar{S}C$ is also symmetric (reciprocal system) if $C_{22}=C_{11}$. Similarly, for $\sigma=\pm i$ ($\gamma=-1$), $S$ always satisfies reciprocity if $C_{22}=-C_{11}$. However, when $\gamma$ is complex, reciprocity and realness of $S=\bar{S}C$ cannot be satisfied with a constant $C$. Since an actual physical system is obviously reciprocal, even when designed for complex $\gamma$, this means that, in this case, $C(\omega)$ is necessarily nonconstant, but rather slowly varying due to other modes, in a way that guarantees reciprocity. In other words, it is not possible to obtain high-$Q$ modes verifying Eq.~(\ref{r-equation}) with complex $\gamma$ without additional modes proximal enough to form a frequency dependent $C(\omega)$. 

\subsection{Geometrical symmetry}

The expression $\sigma_{n}=\pm1$ means that the radiative part of the modes is even or odd, which can be easily obtained using a structure with geometrical (e.g. mirror) symmetry between the two ports~\cite{JoannopoulosJo08-book}. This can explain the increase in transmission previously observed in symmetric structures~\cite{cheron2019broadband}. However, we will later see filter designs where it is preferable for the structure to \emph{not} be symmetric, so the modes themselves are not even or odd, even though their radiative far fields may in fact be (satisfying $\sigma_{n}=\pm1$).

On the other hand, for $\gamma\neq1$, the mode and structure have to be asymmetric anyway. In particular, even-order antimetric SFs ($\gamma=-1$) can be made only from asymmetric structures, as confirmed for example by their known corresponding electric circuit topologies~\cite{dimopoulos2011analog}. However, we explained that good approximate filters can be obtained with $\gamma$ deviating from its optimal value by a phase factor [as long as the background $C(\omega)$ is slowly varying, in contrast to being constant for exact SFs]. Therefore, approximate even-order SFs can also be designed with $\gamma=1$. To highlight this point, we later show implementations of such filters, using \emph{symmetric} structures.

\subsection{Summary}

An $N$-order two-port filter, whose reflection is zero at $N$ real frequencies, obeys $S_{22}=\gamma S_{11}$ ($\gamma=\pm 1$, for symmetric and antimetric) and consists only of modes whose radiation couples to the two ports with the ratios $\sigma_{n}=\pm\sqrt{\gamma}$. To design standard filters, these $\sigma_{n}$ ratios must alternate sign among consecutive resonances, with complex frequencies matching the ``textbook'' filter poles~\cite{dimopoulos2011analog}, and a roughly constant background scattering $C$, appropriate for the desired filter type, must be established [Eqs.~(\ref{eq:condition-a}--\ref{eq:condition-d})]. With other choices of complex pole values or $\sigma_n$-sign orders, one can also design nonstandard filter spectra (see the examples in \figref{Fig-C-background} and Appendix~\ref{app_nonstandard_spectra}). Approximate filters can also be designed with complex unitary $\gamma$. Once the QNMT design objectives (constant background transmission $|C_{21}|$, eigenfrequencies $\omega_{n}$, and port-coupling ratios $\sigma_{n}$) have been determined for the desired filter profile, an implementing physical structure can be found using adequate optimization or nonlinear-solver tools to force the structure to satisfy these objectives. Details on this optimization procedure are given in section~\ref{sec:design_optimization}. It is also shown there that, for the \emph{same} filter design objectives, \emph{different} optimal structures can be found with the \emph{same} desired spectral response, up to small errors arising from other modes outside the bandwidth of interest, leading to a nonconstant $C(\omega)$.

\section{APPLICATION TO MICROWAVE METAFURFACE FILTERS}
\label{sec_fss}

The analytical criteria we have presented in this article give a direct pathway to precisely design high-order two-port filters in all kinds of wave physics (acoustics, photonics, quantum, etc.). As a demonstration, we apply our method to microwave metasurface filters. It is important to clarify up front that, in all examples presented hereafter, we do not use any topology optimization algorithms to determine the structures (although our method can be combined with those, in principle). Instead, for each desired filter response, we choose a fixed topology expected to give \emph{roughly qualitatively} the desired spectral shape (bandpass versus bandstop, number of resonances, etc.) by using physical intuition, which is based on circuit-theory principles and sometimes also on QNMT itself to devise low-$Q$ pole configurations generating the required background scattering $C$ (see, e.g., Ref.~[\citealp{QNMT}, Sec. IV]). The chosen topology for each metasurface has few unknown physical parameters (geometric feature sizes and dielectric permittivities), which are then optimally identified by simply applying a multivariable solver to the nonlinear system of equations for the filter conditions derived in section~\ref{sec-filter-design} to \emph{precisely quantitatively} match the desired SF. This rather ``traditional'' approach leads to rapid computational design, as physics and analytics have already been used to facilitate the job of the optimizer.

For comparison, all filters designed in this article have specifications: center frequency $f_\mathrm{c}=\omega_\mathrm{c}/2\pi=10$ GHz, passband ripple of at most 0.25 dB, and stopband attenuation of at least 25 dB; so only the filter type and bandwidth may differ. For standard filters (all except for Sec.~\ref{subsec_doublelayer}), we easily calculate the desired ideal ``textbook'' poles $\omega_n^{\mathrm{opt}}$ via the \textsc{matlab}$^{\text{\textregistered}}$~\cite{MATLAB_R2019a_u4} ``Signal Processing Toolbox'' functions ``butter,'' ``cheby1,'' ``cheby2,'' and ``ellip,'' while their corresponding $\sigma_n^{\mathrm{opt}}$ are given by the appropriate expressions from Eqs.~(\ref{eq:condition-a})--(\ref{eq:condition-d}). All ideal-filter spectra in \figsref{Fig-double-background}--\ref{Fig-optimizations} (dashed lines) are computed using Eq.~(\ref{S_TC}) with these ideal parameters $\{\omega_n^{\mathrm{opt}},\sigma_n^{\mathrm{opt}},C^{\mathrm{opt}}\}$ (including the corresponding negative modes).

For all microwave metasurfaces: we use a square periodic lattice of period $a$ (with its principal axes along $\hat{x}, \hat{y}$), the metallic material is taken as perfect electric conductor (PEC) with thickness $18\textrm{\ensuremath{\mu}m}$, and the tiny volume of any etched out metal (e.g., inside slits) is taken simply as air. The thickness $18\textrm{\ensuremath{\mu}m}$ corresponds to $0.5$ oz copper, whose finite conductivity has at these frequencies only a small attenuation effect, which is known to get worse as the filter bandwidth decreases~\cite{haus1984waves}, as also demonstrated in the examples later. The \textsc{comsol} Multiphysics$^{\text{\textregistered}}$~\citep{comsol} finite-element software is used (with mesh resolution fine enough to ensure accuracy for the desired spectral features) to carry out the numerical computation of the eigenmodes $\{\omega_n,\sigma_n\}$ during our QNMT-based design, as well as of the ``exact'' frequency-domain response $S(\omega)$ for planewave excitation of the final optimized structures. Specifically, for plane-wave ports $(\vect{E}_p,\vect{H}_p)$, the QNM-to-port couplings are evaluated from the \textsc{comsol} non-normalized eigenfields $(\vect{E}_n,\vect{H}_n)$ as
\begin{equation}
D_{pn}\propto\int_{p}\left(\vect{E}_{p}^{*}\times\vect{H}_n+\vect{E}_n\times\vect{H}_{p}^{*}\right)\;\cdot d\mathbf{S}
\label{D-overlap-integral}
\end{equation}
at the two ($p=1,2$) external port boundaries of the metasurface ($d\mathbf{S}$ points outwards), and then $\sigma_n=D_{2n}/D_{1n}$ (independent of the QNM scaling amplitude). More details regarding the finite-element computations (especially regarding low-$Q$-modes) can be found in Ref.~[\citealp{QNMT}, Appendix~F]. In the Supplemental Material~\cite{supinfo}, we provide tables with the calculated QNMs for every metasurface presented.

\subsection{\!\!\!\!Polarization-preserving transmissive metasurfaces}
\label{sec_fss_transmissive}

In this section, the two-port metasurface filters we design are for transmission of a normally incident plane wave through the metasurface. We choose the period $a$ small enough for filter operation (centered around $f_\mathrm{c}$) below the first diffraction cutoff ($f_\mathrm{cut}=c/a$ at normal incidence), so only transmission and reflection need be considered. The metasurface topologies have planar $p4mm$ ($90^\circ$-rotational plus 4-mirror) symmetry, so the response for normal incidence is independent of the polarization $\hat{e}$, no cross-polarization coupling occurs, and thus indeed only two ports are needed. In this scenario, Eq.~(\ref{D-overlap-integral}) simplifies to $D_{pn}\propto\int_{p}E_{ne} dS$, where $E_{ne}$ is the $\hat{e}$ component of the $n$-QNM electric field. To demonstrate the generality of our design method, we obtain all types of bandpass and bandstop transmission SFs with different orders and bandwidths. In particular, we show SFs with different $|C_{21}|$ values corresponding to all four cases (a)--(d) discussed earlier, as it is instructive for the reader to understand in each case the physical intuition for choosing the appropriate metasurface topology and the required phase relation between $\sigma_n$ and $C$ [Eqs.~(\ref{eq:condition-b},\ref{eq:condition-c})].

Normal incidence is chosen here for simplicity. The angle dependence of the designed filters' response is beyond the scope of this paper, which solely aims to demonstrate the two-port design method. In principle, one can add further constraints to also minimize the performance drop-off away from normal incidence to achieve angle independence or one can apply this method directly for a nonzero angle to design a precise filter at non-normal incidence (in which case, fixed-angle QNMs should be used~\cite{Gras2019angle,QNMT}). Still, for the sake of completeness, in Appendix~\ref{app_angle_dependence}, we show some comparative results, which reinforce the intuition that, among different parameter sets giving the same normal-incidence filter, metasurfaces with smaller period $a$ (by using high permittivities) tend to maintain their performance at an angle better.

\subsubsection{Second-order bandpass filter---circuit model}
\label{subsec_doublelayer}

\begin{figure}
\includegraphics[width=\columnwidth,keepaspectratio]{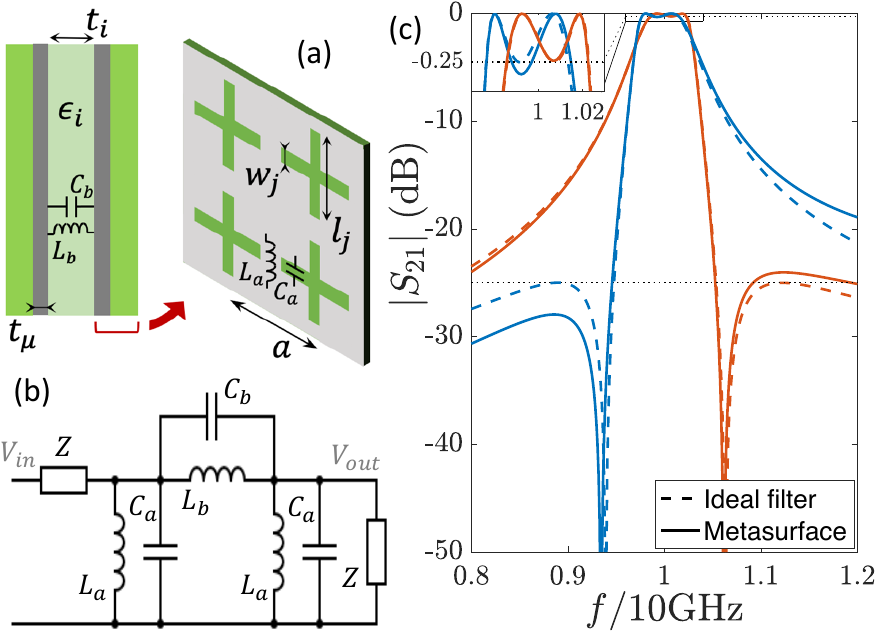} \caption{(a) Symmetric metasurface for a second-order bandpass filter centered at 10 GHz with a single transmission zero, designed for 0.25 dB passband ripple and 25 dB stopband attenuation (black dashed lines). (b) Equivalent circuit model. The coupling $L_{b}C_{b}$ gives the transmission zero. (c) Transmission spectrum of two optimized symmetric structures with a zero respectively on the left and on the right of their transmission peaks. Parameters for left zero are $a$ = 6mm, $w_{1,2}/a=0.0479$, $l_{1,2}/a=0.846$, $t_{1,3}/a=0.493$, $t_{2}/a=0.257$, $\epsilon_{1,3}=1.43$, $\epsilon_{2}=14.51$. Parameters for right zero are $a$ = 9.34mm, $w_{1,2}/a=0.00877$, $l_{1,2}/a=0.905$, $t_{1,3}/a=0.0237$, $t_{2}/a=0.0966$, $\epsilon_{1,3}=4.12$, $\epsilon_{2}=3.80$.}
\label{Fig-double-background} 
\end{figure}

We start by studying a simple \emph{symmetric} second-order metasurface, in order to build some physical intuition on how a particular structural topology can be modeled by an effective circuit, to relate this circuit to the QNMT, and to derive design guidelines for transmission-zero placement. The metasurface, shown in \figref{Fig-double-background}(a), is formed by two planar metallic sheets sandwiched between three uniform dielectric layers. A square array (with period $a$) of narrow crosslike apertures is etched in each metallic sheet, so that the centers of the crosses are the same for all patterned sheets. Each aperture array creates a resonance, which can be modeled in the subwavelength limit ($a\ll\lambda$) as an effective shunt parallel $L_{a}C_{a}$ ($\equiv1/\omega_{a}^{2}$) to a plane wave incoming from free space with impedance $Z$. The inductance $L_{a}$ originates from the current flowing around the edge of the aperture, while the capacitance $C_{a}$ comes from the opposite-charge accumulation across facing sides of this narrow gap [see Fig.~\ref{Fig-double-background}(a)]. The connected topology of the metallic sheets represents a short circuit to an incident plane wave at long wavelengths (shunt $L_{a}$), leading to no transmission at zero frequency. Moreover, a longitudinal inductance $L_b$ couples the apertures on the two metallic sheets, corresponding both to first-order transmission-line effects of the thin dielectric layer and to the direct mutual inductance between the apertures. Finally, capacitance $C_{b}$ builds up between the two metallic sheets  [see Fig.~\ref{Fig-double-background}(a)], which is an interesting feature that has an important consequence: it leads to the emergence, in series with the path of incident-wave propagation (longitudinally), of a parallel-resonant $L_{b}C_{b}$,  which becomes an open circuit at the frequency $\omega_{b}=1/\sqrt{L_{b}C_{b}}$, thus leading to a zero in the transmission function. The final equivalent-circuit model is given in \figref{Fig-double-background}(b), corresponding to a passband filter with a finite-frequency zero.

The transmission spectrum can be computed through $S_{21}=2V_{out}/V_{in}$~\cite{dimopoulos2011analog}, and with $y_{j}=Z\left(1/\omega L_{j}-\omega C_{j}\right)$ for $j=a,b$ we obtain 
\begin{equation}
S_{21}(\omega)=\frac{2iy_{b}}{(1+iy_{a})(1+i(y_{a}+2y_{b}))}.
\label{transmission-circuit}
\end{equation}
This clearly shows transmission zeros at $\pm\omega_{b}$, and also at $\omega=0$, $\omega\rightarrow\infty$ (bandpass behavior). Denoting the loss rates $\Gamma_{j}=1/(2ZC_{j})$, the denominator shows the system poles at $\pm\Omega_1-i\Gamma_1$ and $\pm\Omega_2-i\Gamma_2$, where $\Gamma_{1}=\Gamma_{a}$, $\Omega_{1}\approx\omega_{a}$, $\Gamma_{2}=1/\left(1/\Gamma_{a}+2/\Gamma_{b}\right)$, $\Omega_{2} \approx \Gamma_{2} \left(\omega_{a}/\Gamma_{a}+2\omega_{b}/\Gamma_{b}\right)$. One system resonance is identical to the single-sheet resonance, while the second is also affected by the intersheet couplings: it is always narrower ($\Gamma_{2}<\Gamma_{1}$), and $\Omega_{2}\gtrless\Omega_{1}$ if $\omega_{b}\gtrless\omega_{a}$.

When $\omega$ is close to the positive resonances, the RWA $y_{j}\approx(\omega_{j}-\omega)/\Gamma_{j}$ effectively drops the negative resonances. Then, a partial-fraction expansion of Eq.~(\ref{transmission-circuit}) can be obtained:
\begin{equation}
S_{21}(\omega)\approx\frac{i\Gamma_{1}}{\omega-\left(\Omega_{1}-i\Gamma_{1}\right)}-\frac{i\Gamma_{2}}{\omega-\left(\Omega_{2}-i\Gamma_{2}\right)}.\label{transmission-circuit2}
\end{equation}
This is identical to the QNMT result in Eq.~(\ref{S_TC}) with $\sigma=(1,-1)$ and $C=-I$. 

As mentioned earlier, it is the two \emph{different} values for the decay rates $\Gamma_{1,2}$ (of these two opposite-symmetry modes with fully reflective background) that lead to a transmission zero outside the resonant peaks, which is usually not pointed out in typical CMT models for lossless systems~\cite{suh2004temporal,li2010coupled,Li2020}. We saw from Eq.~(\ref{eq:zero_for2modes}) that this zero $\omega_{b}$ appears on the side of the resonance with the smallest loss rate, which is $\Gamma_{2}$ in this case, so $\omega_{b} < \Omega_{2} < \Omega_{1} = \omega_{a} \Leftrightarrow L_{a} C_{a}<L_{b}C_{b}$ or the opposite order. Therefore, we have a recipe to design the zero for this metasurface, based on the equivalent circuit elements. To translate those to physical structural parameters, we observe that, in the quasistatic limit, $L_{b}\propto t_{2}$ (see Appendix~\ref{app_mutual_inductance}) and $C_{b}\propto\epsilon_{2}/t_{2}$, where $t_{2}$ is the small separation between the two metallic sheets and $\epsilon_{2}$ is the dielectric constant of the separating layer. This means that $L_{b}C_{b}\propto\epsilon_{2}$, which does not depend on $t_{2}$ to first order. On the other hand, $L_{a}C_{a}$ not depend on $t_{2}$ also, but it has a weighted dependence on $\epsilon_{1}$ and $\epsilon_{2}$. The location of the transmission zero relative to the poles then mainly depends on the ratio of permittivities of the two dielectric materials. In particular, a zero at a frequency below the poles is obtained using a large $\epsilon_{2}/\epsilon_{1}>1$.

We can also use Eq.~(\ref{transmission-circuit2}) to compute the full-transmission frequencies ($|S_{21}(\omega_{t})|=1$):
\begin{equation}
\omega_{t}=\frac{\Omega_{1}+\Omega_{2}}{2}\pm\sqrt{\left(\frac{\Omega_{1}-\Omega_{2}}{2}\right)^{2}-\Gamma_{1}\Gamma_{2}}.\label{peak-frequency}
\end{equation}
We see that there are two full-transmission maxima between $\Omega_{1}$ and $\Omega_{2}$, as long as the eigenfrequencies are well separated ($\left|\Omega_{1}-\Omega_{2}\right| > 2\sqrt{\Gamma_{1}\Gamma_{2}}$).

We can now use the QNMT to design second-order bandpass filters with a transmission zero either on the right or on the left of the transmission peaks. These are nonstandard spectra and, in both cases, we numerically find the two complex eigenfrequencies $\omega_n^{\mathrm{opt}}$ for which Eq.~(\ref{transmission-circuit2}) gives the filter specifications stated at this section's introduction (0.25 dB passband, 25 dB stopband ripples) and a 3 dB bandwidth of $6\%$ centered around 10 GHz. Then, we use the multivariable nonlinear-equation solver to find the structural parameters that will make the eigenmodes $\omega_n$ of the metasurface of \figref{Fig-double-background}(a) match those desired eigenfrequencies $\omega_n^{\mathrm{opt}}$ and with port-coupling ratios $\sigma=\pm\left(1,-1\right)$. Results for optimized structures are shown in \figref{Fig-double-background}(c). We note that, as expected, the structure with a transmission zero at smaller frequencies has a larger dielectric constant for the inside layer. We also see that the shapes of the transmission spectra deviate somewhat from the ideal filters. This is mainly due to higher-frequency resonances that affect the scattering matrix (acting as a background $C$) and make it different from the two-poles approximation of Eq.~(\ref{transmission-circuit2}), leading to slight reduction of transmission at low frequencies and increase at higher ones. 

Finally, if we wanted to design the structure using directly the circuit model, we would need to compute the circuit elements $L_{a,b},C_{a,b}$ corresponding to the physical metasurface. This typically requires fitting Eq.~(\ref{transmission-circuit}) to the actual spectral response, which is not efficient for design optimization due to the large number of direct simulations required to locate and accurately fit the sharp spectral features. This is exacerbated by errors introduced by the aforementioned higher-order resonances not encompassed by the circuit model.

\subsubsection{Third-order bandpass [case (a)] filters}
\label{subsub:third-order-bandpass}

Using the QNMT design method, as well as guidance from the previous two-pole bandpass structure, we now design all four SF types mentioned in section~\ref{sec-filter-design} for third-order bandpass filters. We saw that all odd-order bandpass SFs have $C_{21}=0$ [case (a)], so Eq.~(\ref{eq:condition-a}) requires port-coupling coefficients with ratios $\sigma^{\mathrm{opt}}\propto(1,-1,1)$ for the three modes.

\begin{figure}
\includegraphics[width=1\columnwidth,keepaspectratio]{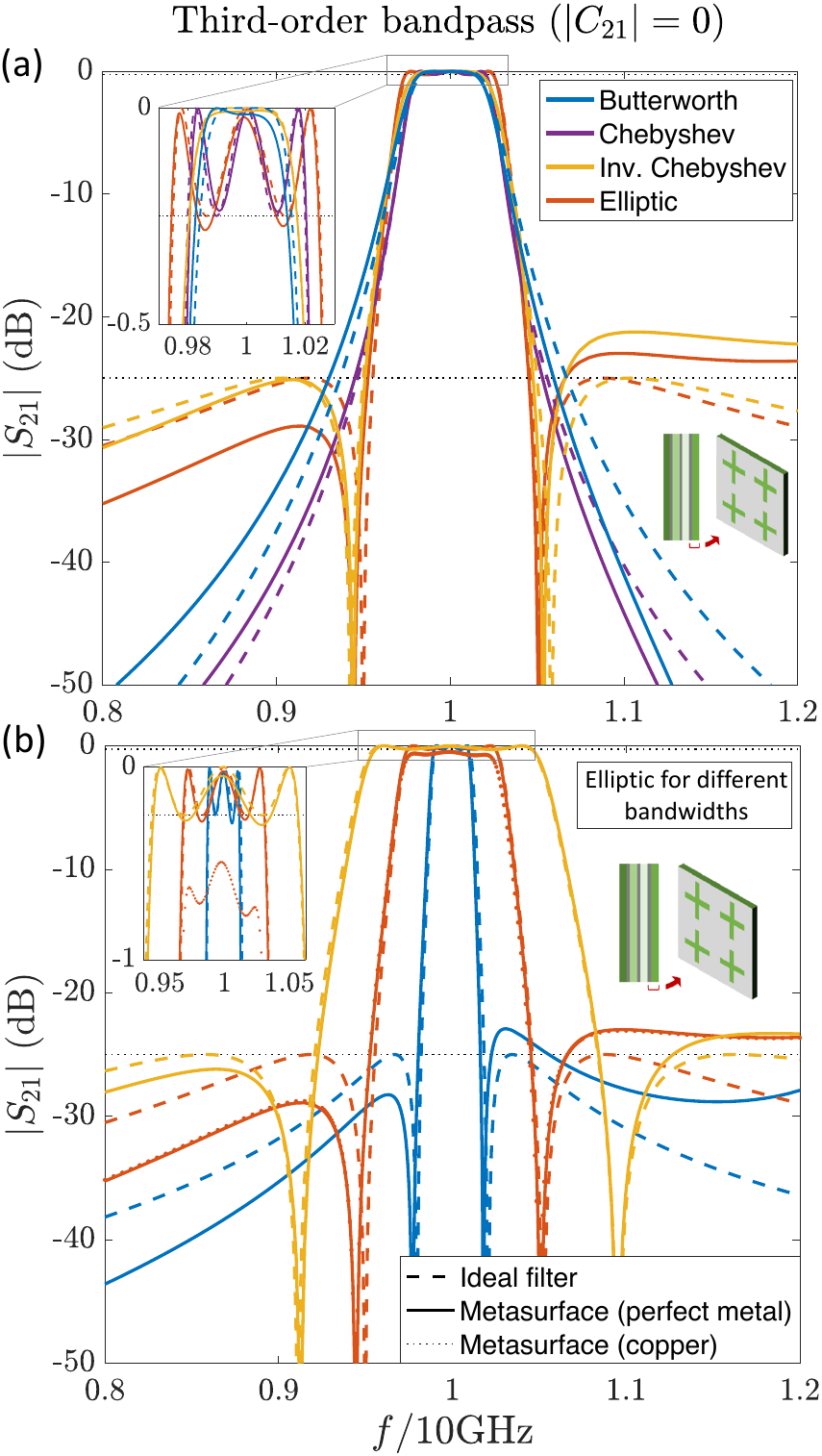} \caption{Optimized third-order bandpass filters (a) of different types with same bandwidth and (b) elliptic only for different bandwidths. We use the same structure as in \figref{Fig-double-background} but with three metallic sheets and four dielectric layers. Physical parameters and 3dB bandwidths are provided in Table~\ref{grid3_params}. We notice good agreement of lossless structures (solid lines) with ideal filters (dashed lines), except for small deviations mainly due to effects from high-order modes. Copper losses (dotted line) reduce peak transmission while preserving the filter's shape.}
\label{Fig-grid3} 
\end{figure}

To implement these filters, we use a structure with the same unit-cell topology as in \figref{Fig-double-background}(a), but with three metallic sheets and four dielectric layers. Based on the insight gained in section~\ref{subsec_doublelayer} from the effective circuit model, we realize that each of the inside layers will create a longitudinal parallel $L_{j}C_{j}\propto\epsilon_{j}$ resonance, which will cause a transmission zero $\propto1/\sqrt{\epsilon_{j}}$. For the inverse Chebyshev and elliptic filters, two \emph{distinct} zeros are required. Therefore, we need \emph{different} dielectric constants $\epsilon_{j}$ for the inside layers. This means that the physical structure for these filter types must \emph{not} be symmetric, so their modes are not even or odd, even if their radiative tails are [$\sigma=\pm(1,-1,1)$]. On the other hand, Butterworth and Chebyshev filters do not have real zeros, so there we can choose a symmetric structure, which simplifies the optimization problem, as only eigenfrequencies need to be matched (in the correct order of modal symmetry). However, for these ``zeroless'' SFs, the challenge with the chosen metasurface topology is to push away from our bandwidth the unavoidable zero that will arise from the inside layers. The simplest way to accomplish this is to look for solutions where these layers are thick enough that the higher-order dependence of the longitudinal parallel $L_{j}C_{j}$ on $t_{j}$ moves the zero to sufficiently high frequencies. Different topologies could also be devised that eliminate either the mutual inductance or capacitance between sheets.

Again, for each filter, by optimizing the structural topology, we force its three complex resonant frequencies $\omega_{n}$ and their corresponding $\sigma_{n}$ to match the desired values $\{\omega_n^\mathrm{opt},\sigma_n^\mathrm{opt}\}$. Transmissions of the optimized metasurfaces that implement the four filter types with approximately 5--6\% 3 dB bandwidth are shown in Fig.~\ref{Fig-grid3}(a), while in Fig.~\ref{Fig-grid3}(b) only for elliptic filters with varying bandwidth (approximately 2--10\%). We note good agreement with the ideal filters, except for small discrepancies again due to effects from higher-order modes and to small errors in the values of optimized resonances. Note that, indeed, Butterworth and Chebyshev filters require thick inside dielectric layers to move the zero away and that smaller bandwidths need higher $\epsilon_{j}$ to increase the modal $Q$s. Moreover, we test the effect of metal (here, copper) losses on the 6\%-bandwidth elliptic filter and find that it mainly just reduces the values of the transmission peaks (only by approximately $0.5$ dB at 10 GHz operation).

\subsubsection{Third-order elliptic bandstop [case (d)] filter}

\begin{figure*}
\includegraphics[width=\textwidth,keepaspectratio]{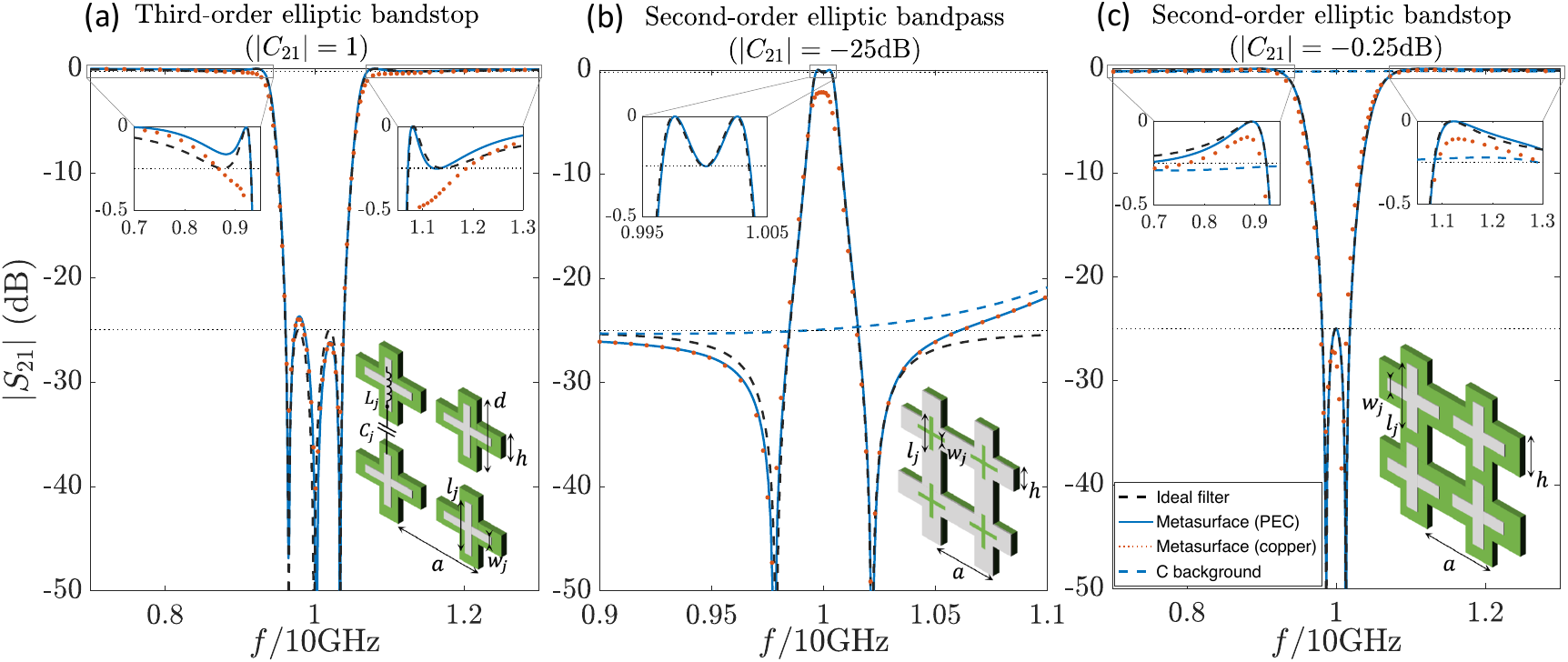} \caption{(a) Third-order elliptic bandstop filter. The structure has three metallic-cross arrays and four dielectric layers with parameters: $a=17.05$mm, $h/a=0.03$, $d/a=0.619$, $w/a=(1.53,3.73,1.61)\times10^{-3}$, $l/a=(0.558,0.589,0.524)$, $t/a=(0.166,0.358,0.441,0.183)$, $\epsilon=(4.76,3.22,4.05,4.50)$. (b) Second-order elliptic bandpass filter. The symmetric structure has two metallic sheets and three dielectric layers with parameters: $a=17.571$mm, $h/a=0.456$, $w_{1,2}/a=6.08 \times 10^{-3}$, $l_{1,2}/a=0.4355$, $t_{1,3}/a=0.3072$, $t_{2}/a=0.3169$, $\epsilon_{1,3}=3.82$, $\epsilon_{2}=1.893$. (c) Second-order elliptic bandstop filter. The symmetric structure has two metallic-cross arrays and three dielectric layers with parameters: $a=18.66$mm, $h/a=0.181$, $w_{1,2}/a=2.74\times10^{-3}$, $l_{1,2}/a=0.514$, $t_{1,3}/a=0.204$, $t_{2}/a=0.332$, $\epsilon_{1,3}=1.60$, $\epsilon_{2}=3.10$. All filters satisfy quite well the marked requirements (black dashed lines) and agree with the ideal filters.}
\label{Fig-structures-background-C} 
\end{figure*}

In order to design a third-order \emph{bandstop} filter, we now need to achieve a full-transmission background $|C_{21}|=1$ [case (d)], and then Eq.~(\ref{eq:condition-d}) dictates again three QNMs with $\sigma^{\mathrm{opt}}\propto(1,-1,1)$. As reviewed in section~\ref{sec-2port-general}, the background-$C$ design can generally be understood using the system low-$Q$ modes and, in particular, a fully transmissive $C$ can be achieved by a mode with infinite radiative rate, which effectively models free space~\cite{QNMT}. Thus, as expected, we need a physical structure with a very small effective index (about $1$), while still able to support the required high-$Q$ resonances. Moreover, since we want full transmission at zero frequency, the metallic components should now not have a fully connected topology. Therefore, relying on the principle of duality, we choose, in place of each metallic sheet with cross apertures, an array of nonconnected thin metallic crosses. These are supported by dielectric crosses, also nonconnected to minimize the total effective index. The structure is shown as an inset in Fig.~\ref{Fig-structures-background-C}(a). Its effective circuit model now sees each array of crosses as a shunt series-$LC$ resonance, where $L$ is the inductance of the cross wires and $C$ is the capacitance across adjacent crosses within each array [see Fig.~\ref{Fig-structures-background-C}(a)]. Then, the couplings between arrays are effectively longitudinal parallel $LC$, where $C$ is the capacitance across facing (cocentric) crosses and $L$ is the first-order transmission-line model of propagation through free space, but also includes the small contribution (a large in-parallel value; see Appendix~\ref{app_mutual_inductance}) of the mutual inductance between facing crosses. This circuit can implement the desired SF, where each shunt series $LC$ or longitudinal parallel $LC$ can directly impose one of the required three distinct transmission zeros.

An example of an optimized structure with a third-order elliptic bandstop response of $11.7$\%  3dB bandwidth is shown in Fig.~\ref{Fig-structures-background-C}(a). We note again the very good agreement compared to the ideal filter. Note that, in duality to the passband filter, the permittivity of one inside dielectric turns out to be \emph{smaller} than the outside layers.

\subsubsection{\!Second-order elliptic bandpass [case (b)] and bandstop [case (c)] filters}

To complete our set of design examples, we now demonstrate second-order elliptic bandpass and bandstop metasurface filters. In this case, we need to design a specific nontrivial background $C$, in particular, $C_{21}$ must be roughly constant within the bandwidth of interest and its amplitude set respectively to the desired stopband minimum attenuation value ($-25$ dB) [case (b)] or passband maximum ripple value ($-0.25$ dB) [case (c)]. Furthermore, the coefficients $\sigma_n$ should respectively satisfy Eq.~(\ref{eq:condition-b}) or Eq.~(\ref{eq:condition-c}). As discussed in section~\ref{sec-filter-design}, for even-order SFs, $\gamma=-1$ and $C$ is a real constant matrix, so $\beta\equiv C_{11}C^*_{21}/\left|C_{11}C_{21}\right|=\pm1$ and $\sigma_{n}^{\mathrm{opt}}=\pm i$, which corresponds to asymmetric  structures, such as the standard circuit topologies of even-order SFs. However, we explained that approximate solutions with a different unitary $\gamma$ are possible (using the RWA) and here we present \emph{symmetric} structures ($\gamma=1$) exhibiting a second-order elliptic filter response within the bandwidth of interest. Since symmetry guarantees $\sigma_n=\pm1$, Eqs.~(\ref{eq:condition-b},\ref{eq:condition-c}) become design objectives for $\beta$, which must respectively match $\beta^\mathrm{opt}=\pm i/\sigma_1$.

For the bandpass filter (with two transmission zeros), we use as a starting point for the structural topology that from Fig.~\ref{Fig-double-background}(a) corresponding to a second-order bandpass filter with only one zero. There, the large metallic sheets led to $C_{21}=0$. In order to increase $\left|C_{21}\right|$ to the small required $-25$ dB around the filter center frequency $\omega_{c}$, we open holes through the entire metasurface, as shown in Fig.~\ref{Fig-structures-background-C}(b), so that some of the incident wave will directly go through without coupling to the high-$Q$ resonances of the crosses. Excluding those two high-$Q$ modes, using QNMT, we calculate $C=-\bar{S}_{\{\omega_n^C,\sigma_n^C\}}$, and it turns out that even-odd pairs of \emph{almost} degenerate low-$Q$ modes below $\omega_\mathrm{c}$ together with higher-order modes lead to a background with a flat small $\left|C_{21}\right|$ and constant $\beta$ over a fairly large frequency range (see the modes in the Supplemental Material~\cite{supinfo}). [Traditionally, one would approximate $C$ by simulating an effective background structure (e.g., where the cross apertures that cause the high-$Q$ resonances are closed), but the result is inaccurate ($-19$ dB instead of $-25$ dB).] The optimization then consists of enforcing the values of the two complex eigenfrequencies, $\left|C_{21}\left(\omega_{c}\right)\right|=-25$ dB and, from Eq.~(\ref{eq:condition-b}), $\beta\left(\omega_{c}\right)=i/\sigma_1$. The transmission of the designed structure is shown in Fig.~\ref{Fig-structures-background-C}(b) and agrees very well with the SF spectrum of $1.1$\% 3 dB bandwidth. It turns out that the modal symmetry is $\sigma=\left(-1,1\right)$, so $\beta\left(\omega_{c}\right)=-i$. Note that, since each metallic sheet is still connected (in a topological sense), the transmission at very long wavelengths will still go to zero.

For the second-order bandstop elliptic filter, we use as a starting point the structural topology from Fig.~\ref{Fig-structures-background-C}(a) for the third-order bandstop filter, but with two metallic-cross arrays sandwiched between three dielectric layers. There, the effective refractive index of the entire metasurface is designed small to get $\left|C_{21}\left(\omega_{\mathrm{c}}\right)\right|\approx1$. In order to decrease $\left|C_{21}\right|$ to the required $-0.25$ dB, we connect the dielectric crosses, as shown in Fig.~\ref{Fig-structures-background-C}(c), to reflect back some of the incident wave. In QNMT terms, an averaging over the metasurface leads to an effective slab of low refractive index, which supports equispaced ``Fabry-Perot'' low-$Q$ modes $\omega_{n}^{C}=n\Omega^C-i\Gamma^C$ \cite{haus1984waves}; the ``Fabry-Perot'' transmission hits $\left|C_{21}\right|=1$ at $\Omega_{n}^{C}$, but is less than 1 and flat between modes $[n\Omega^C,(n+1)\Omega^C]$, with roughly constant $\beta\approx i$ if $n$ is even (and $\beta\approx -i$ if $n$ is odd). For this structure, it turns out that these modes $\omega_{n}^{C}$ have such a large $\Gamma_n^C$ (see the modes in the Supplemental Material~\cite{supinfo}) that it is difficult to accurately find all higher-$n$ modes still contributing to $C\left(\omega_{c}\right)$, as the relevant region of the complex plane is polluted by the branch cut associated with a higher-order diffraction port. Therefore, we instead calculate it indirectly from  $C(\omega_\mathrm{c})=\bar{S}^{-1}(\omega_\mathrm{c})S(\omega_\mathrm{c})$, as explained in section~\ref{subsec:conditions}. [Again, the traditional method of an effective background structure (removing the metallic crosses) gives a noticeably inaccurate estimate of $C$ ($-0.11$~dB instead of $-0.25$~dB).] The two high-$Q$ modes have symmetry $\sigma=\left(-1,1\right)$, consistent with $\beta\left(\omega_{c}\right)=i$ from Eq.~(\ref{eq:condition-c}), and the optimized final structure has  transmission shown in Fig.~\ref{Fig-structures-background-C}(c), matching precisely an elliptic bandstop SF of $10.3$\% 3 dB bandwidth.

\subsection{Polarization-converting reflective metasurfaces}

To demonstrate different port configurations, we now examine plane-wave normal incidence on microwave metasurfaces without planar $p4mm$ symmetry, so the two polarizations couple. To maintain the number of ports at two, which is the scenario of applicability of our design criteria, we consider metasurfaces that still have small enough period $a$ to avoid diffraction ($f_\mathrm{c}<f_\mathrm{cut}=c/a$), but now also have a full metallic backing. In this way, a plane wave of one polarization (port $\hat{x}$) can only be reflected, either onto its own polarization or onto the other polarization (port $\hat{y}$). Note that, although a full $\hat{x}\rightarrow\hat{y}$ polarization conversion may seem like a $90^\circ$ rotation, this is in fact merely a ``polarization reflection'' across a diagonal $\hat{x}\pm\hat{y}$ plane, since reciprocity prevents having an actual $90^\circ$ rotator of any incident polarization, which would require $S_{yx}=-S_{xy}$. For example, when $|S_{xy}|=1$, an incident wave linearly polarized along $\hat{x}\pm\hat{y}$ keeps the same polarization upon reflection from the metasurface. Such metasurfaces are called ``reflective polarization converters''~\cite{Grady2013, Feng2013, Zhang2016, Loncar2018, Karamirad2021} and, since they are one-sided, they have planar ``wallpaper symmetry groups''~\cite{IntTablesCrystal2016}.

\begin{figure}
\includegraphics[width=\columnwidth,keepaspectratio]{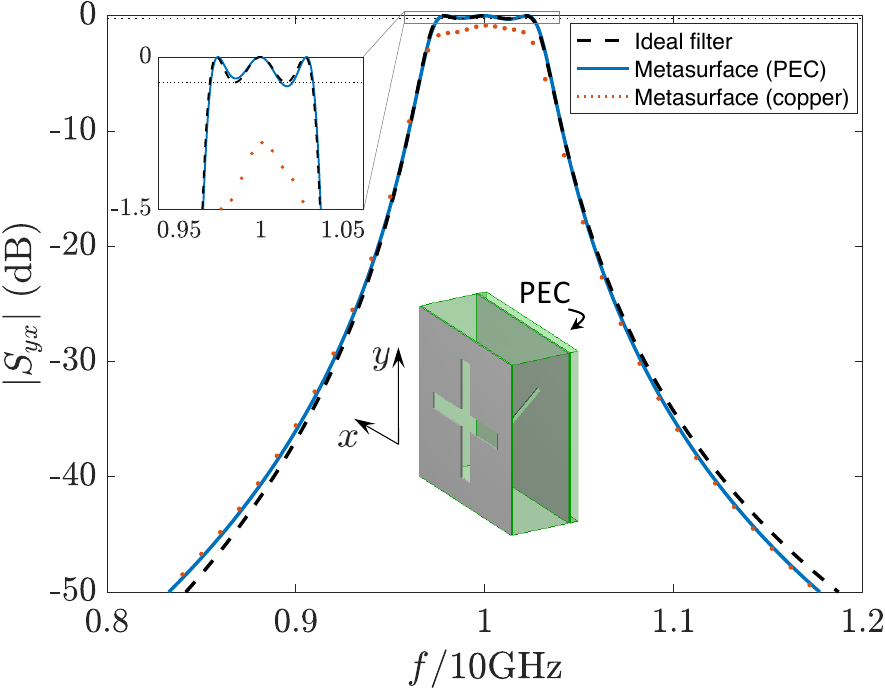} 
\caption{Polarization-converting reflective metasurface with third-order Chebyshev filter response. The symmetric (across the $\hat{x}\pm\hat{y}$ planes) structure has two metallic sheets and two dielectric layers, on top of a back reflector, with parameters: $a=8.138$mm, $w_{1,3}/a=0.1024$, $w_{2}/a=6.035\times10^{-3}$, $l_{1,3}/a=0.7807$, $l_{2}/a=0.6783$, $t_{1}/a=0.5142$, $t_{2}/a=0.0556$, $\epsilon_{1}=7.492$, $\epsilon_{2}=7.789$.}
\label{Fig-pol-rotator}
\end{figure}

For brevity, we only consider a third-order bandpass Chebyshev filter in polarization conversion $S_{yx}$. Namely, an incident plane-wave wide-spectrum pulse will be reflected from the metasurface, keeping the same polarization in the stopband, but having its polarization ``reflected'' with respect to the $\hat{x}+\hat{y}$ axis in the passband. For third-order bandpass, we need $|C_{yx}|=0$ and three resonances, while the Chebyshev shape can be implemented most simply with a symmetric structure with an $\hat{x}+\hat{y}$ symmetry plane [so $\sigma=\pm(1,-1,1)$ automatically, where $\sigma_n=D_{yn}/D_{xn}=\int E_{ny} dS/\int E_{nx} dS$, calculated at the front face of the metasurface]. Using intuition from our transmissive third-order bandpass topologies of Fig.~\ref{Fig-grid3}, we need slit apertures on metallic sheets, where ports (polarizations) $\hat{x}$ and $\hat{y}$ couple respectively only to the resonances of the first and third slits, which are only cross-coupled via the second slit. This is accomplished by the metasurface shown at the inset of Fig.~\ref{Fig-pol-rotator}, respecting both $\hat{x}\pm\hat{y}$ symmetry planes: a symmetric-cross aperture on a front metallic sheet provides two resonances without mixing the polarizations and an additional diagonal slit on a second metallic sheet (between the front sheet and the perfect metal reflector) is the only element that breaks the wallpaper symmetry group $p4mm$ ($90^\circ$ rotational plus 4 mirrors) down to a $c2mm$ ($180^\circ$, including rotation centers off 2 mirrors)~\cite{IntTablesCrystal2016}, so it couples the two front slit resonances.

By optimizing over the structural parameters, we find a set of slit dimensions and dielectric layers' thicknesses and permittivities that forces the QNMs of the structure to match the poles of a 6.3\% 3-dB-bandwidth Chebyshev SF. As seen in \figref{Fig-pol-rotator}, the exact frequency domain simulation gives a spectral response for polarization conversion upon reflection that matches remarkably the desired Chebyshev shape.

\subsection{Diffractive reflective metasurfaces}

As our last application, we consider yet another metasurface two-port configuration. A metal back reflector is again present, but the plane wave now has a frequency $c/2a<f<c/a$ and is incident (with the wave vector in the $\Gamma\mathrm{X}_x$ direction of the $k_x k_y$ Brillouin zone) at an angle $\theta>\mathrm{arcsin}(c/af-1)$, so that the $-1$ diffracted beam appears at the angle $\theta_{-1}=\mathrm{arcsin}(\mathrm{sin}\theta-c/af)$, while all other diffraction orders (spatial harmonics) are evanescent (i.e., outside the light cone). With excitation along $\Gamma\mathrm{X}_x$, a system with $y$-mirror symmetry still decouples the two polarizations, therefore, for one of them incident, only two ports are again present (the $0,-1$ beams). When full conversion from the $0$ to the $-1$ order is achieved, this phenomenon has been called ``perfect anomalous reflection''~\cite{Hessel1975,Tretyakov2017,Radi-Alu2017,Eleftheriades2018,Epstein2019}.

\begin{figure}
\includegraphics[width=\columnwidth,keepaspectratio]{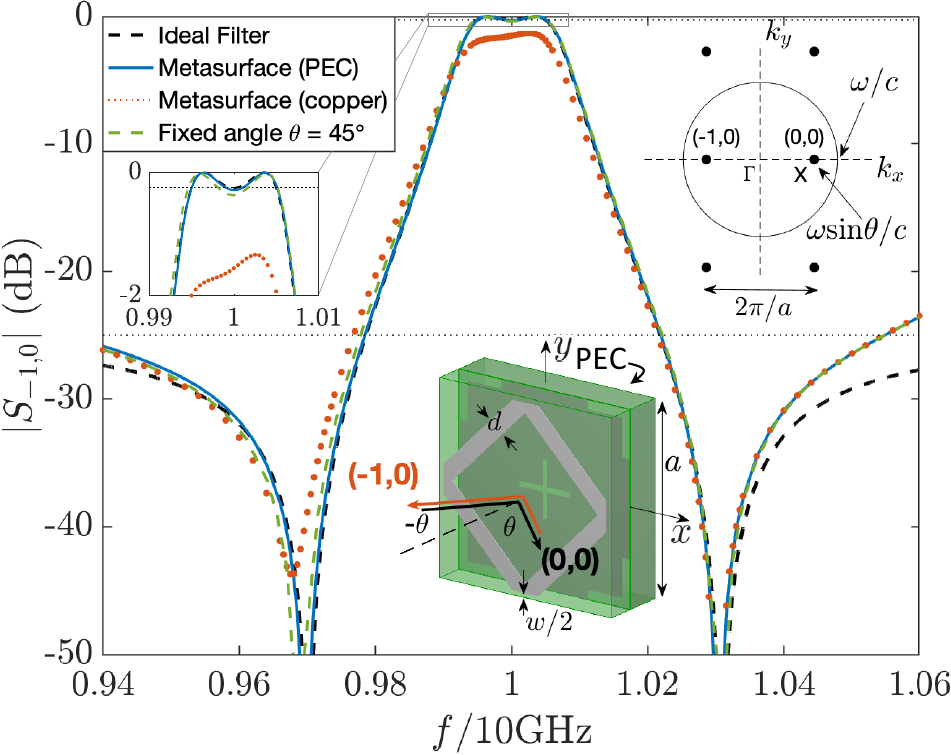} 
\caption{Diffractive reflective metasurface with second-order elliptic filter response. The bottom center inset illustrates the structure unit cell ($a=21.2$ mm), which is stacked (from top to bottom) as follows: square-loop metallic stripe of width $d=0.1354a$ and chamfered corners to get inter-loop gap $w=0.004406a$; dielectric layer $\epsilon=9.8$, $t=0.06804a$; metallic sheet with etched center cross of $w_1=0.009184a$, $l_1=0.2731a$, and corner crosses of $w_2=0.0301a$, $l_2=0.259a$; dielectric layer $\epsilon=2$, $t=0.1455a$; back (PEC) reflector. Within the filter bandwidth, the $\hat{y}$-polarized incoming ``$0$'' (black) wave at angle $\theta$ is ``anomalously'' reflected back into the same direction $-\theta$, corresponding to the ``$-1$'' (red) diffracted beam. (top right inset) Among all Bloch wave vectors (black dots), only the incident ``0'' wave at the Brillouin-zone X edge ($k_{x,0}=\omega\mathrm{sin}\theta/c  =\pi/a$) and the ``$-1$'' diffracted order (at $k_{x,-1}=-\pi/a$) are inside the light cone ($\omega/c$ circle in $k_x k_y$ plane) and are thus propagating ports.}
\label{Fig-diffraction} 
\end{figure}

In this demonstration example, we choose the incidence angle $\theta(f)=\mathrm{arcsin}(c/2af)$, so that the metasurface operation is exactly at the Brillouin-zone edge ($\mathrm{X}_x$ point) at all frequencies. Choosing a fixed Bloch wave vector $k_{0,xy}=(\pi/a,0)$ (instead of a fixed angle) makes the simulations much simpler and also highlights the ``anomalous'' reflection in that, within the filter passband, the obliquely incident wave is reflected back to where it came from [$\theta_{-1}(f)=-\theta(f)$, called ``retroreflection'']. We target $\theta(f_\mathrm{c}=10\mathrm{\;GHz})=45^\circ$, so we need $a=21.2$ mm. The configuration of the ports and of the propagating spatial harmonics are shown as insets of \figref{Fig-diffraction}. For the two-port regarding the $\hat{y}$ polarization, we now wish to design for $S_{-1,0}$ a second-order elliptic bandpass filter of 1.6\% 3 dB bandwidth. Namely, only for a very narrow range of frequencies does anomalous reflection occur (back at $-\theta$); otherwise (in the stopband), the wave is regularly reflected (at $\theta$). Here, Eq.~(\ref{D-overlap-integral}) gives $\sigma_n=D_{-1n}/D_{0n}=\int E_{ny}e^{-i\pi x/a} dS/\int E_{ny}e^{i\pi x/a} dS$. The $\hat{x}$-mirror symmetry ensures that $\sigma=\pm(1,-1)$ for the two high-$Q$ resonances, so we have to achieve the necessary $|C_{-1,0}|=-25$ dB with $\beta^{\mathrm{opt}}=\mp i$, from Eq.~(\ref{eq:condition-b}). We first design this slowly varying (softly diffracting) $C$ with a lattice of metallic stripes, disconnected at the corners by narrow slits to form square loops with chamfered corners and placed on top of a metal-backed dielectric layer. We roughly optimize parameters to get a quite flat $|C_{-1,0}|\approx-25$ dB around $f_\mathrm{c}$ and $\beta=+i$ (so we need $\sigma_1=-1$). Then, on the second metallic sheet, we open two coplanar arrays of dissimilar slits and add a final PEC-backed dielectric layer to form the final ($p4mm$-wallpaper-symmetry) metasurface (see the inset of \figref{Fig-diffraction}). The parameters are then optimized (using $C=\bar{S}^{-1}S$, since the diffraction branch cuts pollute the low-$Q$ region of the complex plane) to give resonances at $\{\omega_n,\sigma_n=(-1,1)\}$. We get a diffraction spectrum, which once again matches the desired filter response (\figref{Fig-diffraction}). We also show the response of the designed structure at a fixed angle $\theta=45^\circ$ (green curve) and we see that it is almost identical to that at fixed $k_x$.

It should be clear that one can also design such ``perfect anomalous reflection'' filters away from $\mathrm{X}_x$ for a different pair of incidence and diffraction angles.

\subsection{Fabrication and tunability}
All the metasurface filters that we have presented are based on a layered topology with metallic sheets sandwiched between dielectric layers. This layered form is chosen because it has the great advantage of allowing easy fabrication. Especially in the cases where patterning is only on the metallic sheets, these metasurfaces can be manufactured even with widespread printed-circuit board (PCB) techniques. In fact, all designed SFs presented in this article used dielectrics with permittivities less than $11.2$, which is roughly the upper limit for low-loss (typically $\text{Al}_2 \text{O}_3$-based, $\mathrm{tan}\delta\lesssim0.0025$) materials compatible with PSBs. Furthermore, the clear separation between metallic sheets allows them to be connected to separate electrodes, where voltage can be applied to potentially tune the permittivity of the intermediate dielectrics, if those are chosen to be tunable materials (liquid crystals, ferroelectrics etc.)~\cite{ahmed2015tunableMat}. Previous attempts at elliptic filters have usually employed topologies with shunt metal paths connecting different metal sections within the metasurface, which hinders both these benefits~\cite{luo2007design, lv2019wide, li2013synthesis}.

\section{DESIGN OPTIMIZATION}
\label{sec:design_optimization}

Device inverse design via optimization is widely accepted to be a challenging task. All possible methods face difficulties, such as objective functions with a plethora of local optima or with bad behavior (e.g., nonanalyticity), slow convergence, violation of constraints, etc. Therefore, to find an appropriate structural topology and a ``good'' optimal solution, it may often take a few attempts, including trying different optimization algorithms and settings, several (random or intuition-guided) initial structures, etc. Similarly, the QNMT-based design method we introduced in this article does not lead to trivial optimization problems. To accelerate the solution of our microwave metasurface SF designs in the previous section, we employed physical intuition (e.g., from circuit theory) to choose the topology and we performed few preliminary computations to determine an arbitrary but reasonable initial structural-parameter set for optimization (e.g., to ensure that the lowest-order slit resonances were used). Here, we present more details regarding our optimization (root-finding) procedure and demonstrate with comparative examples that, for strongly coupled wavelength-sized systems, our QNMT analytical design criteria can be more effective than a direct approach of brute-force optimizing the desired spectral response at a finite set of key frequencies. In particular, we show an example where, using \emph{different} initial structural-parameter sets for the \emph{same} filter design objective, our method leads to \emph{different} optimal structures with the \emph{same} (up to small deviations) desired spectral response within the bandwidth of interest, while the brute-force approach fails to converge to the desired response.

\subsection{Optimization objectives and settings}

The analytical design criteria derived in \secref{sec-filter-design} can be enforced using a root-finding problem to set the QNM parameters $\omega_n\equiv\Omega_n-i\Gamma_n$, $\sigma_n$, and background $C$ (if needed) to the required values $\omega_n^{\mathrm{opt}}\equiv\Omega_n^{\mathrm{opt}}-i\Gamma_n^{\mathrm{opt}}$, $\sigma_n^{\mathrm{opt}}$ [from Eqs.~(\ref{r-equation}) with necessary sign order], and $C^{\mathrm{opt}}$. To precisely match the desired spectrum, the resonant frequencies must converge to their target values in the complex plane with an accuracy of the order of their linewidths, so we rescale our complex-frequency errors as $\delta_\omega=\{(\omega_n-\omega_n^{\mathrm{opt}})/\Gamma_n^{\mathrm{opt}}\}\rightarrow0$. (Note, however, that often the rates $\Gamma_n$ have a slower dependence on structural parameters than the real frequencies $\Omega_n$, so, in the first optimization steps, it might be appropriate to use smaller error weights for the $\delta_\Omega$.) When a geometric scaling law can be used (e.g., for Maxwell's equations) and no dimension must be fixed to a specific value, one can eliminate one real-frequency objective $\Omega_\mathrm{o}\rightarrow\Omega_\mathrm{o}^{\mathrm{opt}}$ by multiplying, after each iteration, all dimensions with $\Omega_\mathrm{o}/\Omega_\mathrm{o}^{\mathrm{opt}}$ or the average $<\Omega_n/\Omega_n^{\mathrm{opt}}>$. For a $N$th-order system, this leads to a system of $2N-1$ real equations for $\{\Omega_n,\Gamma_n\}$. 

When the structure is symmetric with respect to the two ports, so that $\sigma_n=\pm 1$ automatically, it is often advantageous to order the modes according to their desired symmetry order before computing the frequency errors. When there is no symmetry, since an overall phase factor is allowed for all $\sigma_n^{\mathrm{opt}}$, we extract the phase $\sigma_l/|\sigma_l|\equiv e^{i\chi_l}$ for some $l$ and enforce only $\sigma_{n,l}\equiv\sigma_n e^{-i\chi_l}\rightarrow\sigma^{\mathrm{opt}}_{n,l} \equiv\sigma_n^{\mathrm{opt}}/\sigma_l^{\mathrm{opt}}=\pm1$ (for $n=l$ this is simply $|\sigma_l|\rightarrow1$). During optimization, potential modal-frequency crossings can be problematic, especially when numerically calculating derivatives, so the modes should be tracked, for example, using their $\sigma_{n,l}$ values. The errors $\delta_\sigma\rightarrow0$ can be formed in many different ways, but $|\delta_\sigma|$ should ideally be invariant under port swaps $\sigma\rightarrow1/\sigma$; for example, one can choose $\delta_\sigma=(\sigma_{n,l}+1/\sigma_{n,l})/2-\sigma^{\mathrm{opt}}_{n,l}$ [which one could optionally further multiply by a factor $(\sigma_{n,l}-\sigma^{\mathrm{opt}}_{n,l})/(\sigma_{n,l}+\sigma^{\mathrm{opt}}_{n,l})$ to maximize the error for the wrong $\sigma_{n,l}$ sign]. Note that, due to the reciprocity condition, Eq.~(\ref{eq:TC_reciprocity}), the real and imaginary parts of $\sigma_n$ are not completely independent, so there may be ways to reduce the number of target equations.

In the cases where a nonzero $|C_{21}^{\mathrm{opt}}|$ or $|C_{11}^{\mathrm{opt}}|$ is required [e.g., Eq.~(\ref{eq:condition-b}) or (\ref{eq:condition-c})], a phase condition of the form $i\sigma_l\beta=\pm1$ must also be satisfied. Since phase $\chi_l$ was removed from all $\sigma_n$, the $C$ requirements can be written as a combined directive $\delta_C=i e^{i\chi_l}C_{11}(\omega_\mathrm{c})C^*_{21}(\omega_\mathrm{c})/|C^{\mathrm{opt}}_{11}C^{\mathrm{opt}}_{21}|\pm1\rightarrow0$. Note that, when the structure is symmetric ($\chi_l=0$, $C_{11}=C_{22}$), unitarity of $C$ immediately leads to $\mathrm{Im}\{\delta_C\}=0$, so one needs to design only $\mathrm{Re}\{\delta_C\}\rightarrow0$. Moreover, since $C$ must be fairly slowly varying around the filter center frequency $\omega_\mathrm{c}$, one may need to impose additional constraints. This can be done, for example, by minimizing $\delta_C$ also at other frequencies in the bandwidth of interest or some derivatives $d^k C(\omega_\mathrm{c})/d\omega^k$ for $k=1,2,...$. (Note that, when the convenient fitting formula $C=\bar{S}^{-1}S$ is used, approximation errors may lead to small oscillations of $C$ around the high-$Q$ resonances close to $\omega_\mathrm{c}$, in which case it is better to use this formula at a few frequencies outside the filter bandwidth, and interpolate for the $C$ value and derivatives at $\omega_\mathrm{c}$ if needed. In contrast, when calculated directly from QNMT, $C=-\bar{S}_{\{\omega_n^C,\sigma_n^C\}}$ does not have these issues and may be preferable if accurate enough.) In some sense, our method effectively isolates the fast spectral oscillations due to high-$Q$ resonances and applies the common brute-force method only for the slowly varying background $C$.

Our analytical QNMT formulation also allows for alternative objectives instead of $\sigma_n$ and $C$. Equations~(\ref{S_TC}) can be used to write $S_{pq}$ as a rational function and then directly compute its zeros $z^{pq}_m$ and an overall multiplicative factor $A^{pq}$. Those can then be used as alternative variables to be directly set by optimization to the required values for the ideal filter. As mentioned earlier, realness and reciprocity require the zeros $z^{21}_m$ to be either real $(z,-z)$ pairs or complex $(z,-z^*,-z,z^*)$ quadruplets (and $z=0$ is matched with a $z\rightarrow\infty$). Therefore, for a $N$th-order system with total $2N$ zeros, targeting only the independent degrees of freedom gives at most $N$ equations for $\mathrm{Re}\{z^{21}_m\},\mathrm{Im}\{z^{21}_m\}$. For example, for a third-order bandpass elliptic filter, this means directly setting $(\mathrm{Re}\{z^{21}_1\}, \mathrm{Re}\{z^{21}_2\})$ to the two positive real elliptic zeros (while the structural topology can often be chosen to ensure that $z^{21}_0=0$).

To optimize the structure, we pass the error vector $(\delta_\omega,\delta_\sigma,\delta_C)$ [or alternatively $(\delta_\omega,\delta_z,\delta_A)$] into a numerical root-finding routine. We simply use the ``fsolve'' function in \textsc{matlab}~\cite{MATLAB_R2019a_u4}, mostly with the Levenberg--Marquardt algorithm, with or without a Jacobian scaling, typically with numerical derivatives based on central differences. The ``fsolve'' function does not support parameter bounds, but we implement them using a hyperbolic-tangent mapping. (To speed up the initial iterations, one can also use a coarser spatial-discretization mesh, larger numerical-derivative step sizes, and/or different weights on the error vector.)

As a final remark, this optimization setup assumes a rather small number of structural parameters. However, our QNMT design criteria can, in principle, be combined with a full topology-optimization setup with a large number of unknown parameters. Such a formulation requires further research to maximize computational efficiency and is beyond the scope of this paper.

\begin{figure}[!h]
\includegraphics[width=\columnwidth,keepaspectratio]{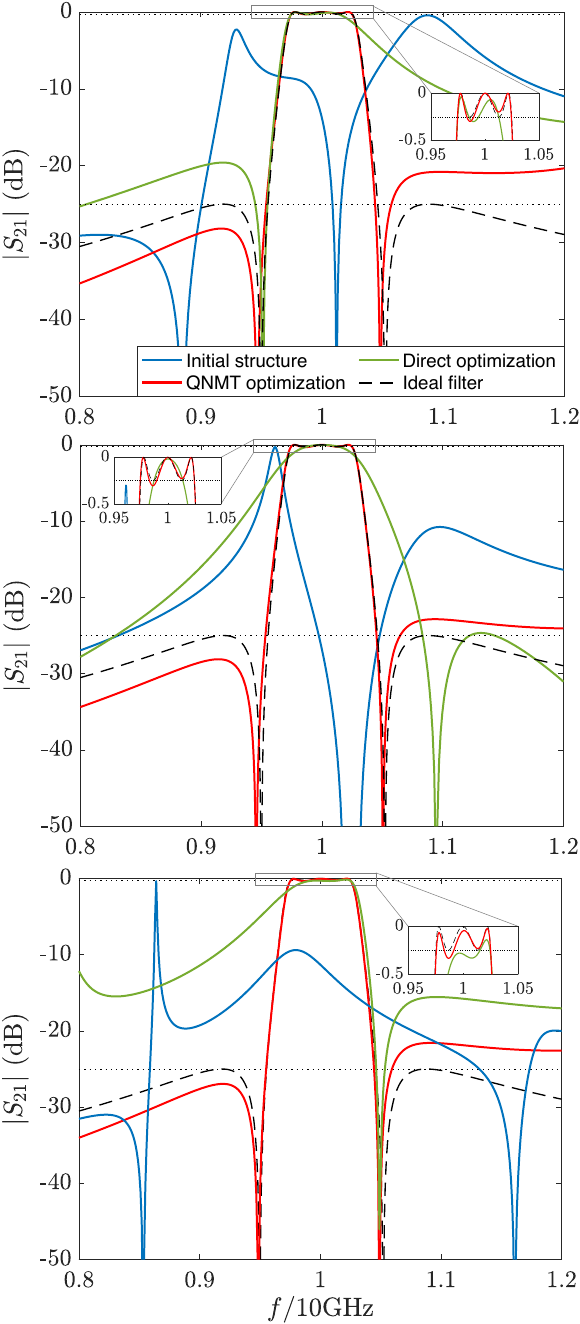} 
\caption{Different optimization solutions for a desired third-order bandpass elliptic SF (black dashed lines). Starting with three different initial structures (blue lines), the solutions obtained with our QNMT method (red lines; structural parameters in Table~\ref{grid3_params}) match the SF, while local optima obtained by directly optimizing the transmission spectrum at few key frequencies (green lines) fail.}
\label{Fig-optimizations} 
\end{figure}

\subsection{Dependence on initial parameter sets and comparison to brute-force optimization}

In order to demonstrate the effectiveness of our QNMT analytical criteria for optimization, we design the third-order bandpass 6\%-bandwidth elliptic response of \figref{Fig-grid3} (dashed red curve), using the same metasurface topology of \figref{Fig-grid3} (inset), but starting with three different sets of initial parameters. The corresponding spectral responses of the initial structures are shown with blue curves in \figref{Fig-optimizations} and can be seen to deviate substantially from the target response. Using dimensional scaling (the fixed 18~$\mu m\ll\lambda,a$ metal thickness has minimal effects, which may only need to be addressed at the very end of optimizations) and $|C_{21}|\approx0$ automatically from the connected-metal metasurface topology, we have 11 real optimization objectives (2 $\delta_\Omega$, 3 $\delta_\Gamma$, 3 $\mathrm{Re}\{\delta_\sigma\}$, 3 $\mathrm{Im}\{\delta_\sigma\}$). We are indeed able in all cases to find different structural parameters (\figref{Fig-optimizations}, red curves) with a transmission spectrum very close to the ideal response. The small stopband discrepancies are due to a slowly varying $C$, whose variations differ for each structure according to each one's resonant content outside this range. Basically, there are many metasurfaces with the same topology that have almost the same spectral response at normal incidence, where each one of those structures can be an optimization solution for some initial point and optimization settings. However, these metasurfaces have different angular responses. In particular, structures with smaller periods tend to be less angle dependent (Appendix~\ref{app_angle_dependence}). The optimization time is obviously highly dependent on the initial structures, the number $P$ of optimized structural parameters, the optimization settings used (e.g., central-difference derivatives lead to $2P+1$ error-vector evaluations per algorithmic iteration), and the termination criteria, but, to give a sense of its order of magnitude, the three designed systems of \figref{Fig-optimizations} took respectively about $150$, 260 ($P=12$), and 300 ($P=9$) evaluations (Maxwell eigenvalue solves).

We now compare to a direct optimization approach based on directly computing the transmission using a frequency-domain solver. To design structures, we minimize the mean-square transmission error compared to the ideal third-order bandpass elliptic response at 9 key frequencies (3~transmission peaks, 2~passband frequencies with $-0.25$~dB transmission, 2~transmission zeros, and 2~stopband frequencies with $-25$~dB transmission).  After trying a variety of error-vector formulations and optimization settings, the best obtained local optima for the three initial structures are shown in \figref{Fig-optimizations} (green curves). We see that they substantially fail to match the desired response. 

There are initial parameter sets for which neither design method manages to converge to a solution. However, the above comparison supports our original claim that, for highly resonant spectra, our QNMT method achieves good solutions for more initial parameter sets compared to brute-force transmission optimization, which tends to converge to poor local optima.

\section{CONCLUSIONS}

We have presented a systematic method using eigensolvers for designing symmetric or antimetric filters [such as standard filters or other useful transmission and reflection spectra (e.g., Appendix~\ref{app_nonstandard_spectra})], especially those with multiple finite real zeros, allowing ultracompact two-port devices with spatially overlapping resonances (unlike previous circuit-theory or CMT approaches). It is based on a non-normalized QNM expansion of the system scattering matrix $S$ and entails identifying the necessary background response $C$, the exact complex eigenfrequencies $\omega_{n}$ of these modes, and the values of the ratios $\sigma_{n}$ with which these modes must couple to the system ports, to achieve the desired scattering frequency profile. An efficient optimization procedure is then applied to determine the structural parameters (geometry and materials) that meet these criteria. We have demonstrated the method for microwave planar metasurface filters, with two-port configurations involving same-polarization transmission, cross-polarization reflection, or diffractive reflection, for all standard amplitude-filter types (especially the most challenging, elliptic), for both bandpass and bandstop behaviors, and for a variety of frequency bandwidths.

Our design method is demonstrated for microwave metasurfaces, but it can also be used for resonant systems with any qualitatively similar wave physics, such as mechanical, acoustic, photonic, or quantum-scattering filters. In our examples, we use fixed topologies, guided by general physical intuition, and then apply a simple multivariable equation solver to obtain a small set of structural parameters. In principle, our conditions can also be combined with various large-scale topology-optimization algorithms (where every ``point'' in space is a degree of freedom) and solver methods~\cite{jensen2011topology}. While we provide analytical criteria for two-port scattering systems satisfying $S_{22}(\omega)=e^{i\varphi}S_{11}(\omega)$ and we focus our examples on the subset of amplitude standard filters, our design process can be used for \emph{any} desired scattering spectrum, by fitting it to QNMT to extract the corresponding optimization objectives $\{\omega_n^{\mathrm{opt}},\sigma_{n}^{\mathrm{opt}},C^{\mathrm{opt}}\}$. Moreover, it should be clear that the accurate QNMT prediction of the time delay~\cite{QNMT} also makes the theory applicable to design phase filters, such as all-pass delay filters~\cite{tsilipakos2018phaseMS} (useful also for metalenses~\cite{capasso2020metalens}). Our approach is likely most suited to and advantageous for fast-varying spectra related to sharp resonances, but nothing really prevents its applicability to broadband systems. This design method assumes lossless two-ports, so it is best done ignoring all losses and is thus limited to systems with only small absorption and weak additional radiation channels. Extension to more than two ports should be possible, since a spectral response $S_{pn}(\omega)$ for any number of ports could be reduced via QNMT to a set of required $\{\omega_n^{\mathrm{opt}},D_{pn}^{\mathrm{opt}},C^{\mathrm{opt}}\}$. For example, in a multiport scenario where full conversion between only two ports, $1,2$, is required, additional conditions $D_{pn}\approx0$ for all other ports $p\neq1,2$ might suffice. Our QNMT is more accurate for port modes with frequency-independent transverse profiles (our metasurfaces used planewave ports), however, the design method could also be extended to most other common ports (e.g., wave guides, Gaussian beams, fixed-angle diffracted waves), when their modal profiles have a \emph{slower} frequency variation than the desired spectral response, potentially by approximating $D_{pn}(\omega)\approx D_{pn}(\Omega_n)$.

\begin{acknowledgments}
This work is supported in part by the U.S. Army Research Office through the Institute for Soldier Nanotechnologies at MIT under Grant no. W911NF-18-2-0048, by the Simons Foundation collaboration on Extreme Wave Phenomena, and by Lockheed Martin Corporation under Grant no. RPP2016-005.
\end{acknowledgments}

\appendix

\section{TRANSFER-MATRIX FORMALISM}
\label{app_transfer_matrix}
For a two-ports system as in \figref{Fig-2port}, the (forward) transfer $T$ matrix is defined as:
\begin{equation}
    \begin{pmatrix}s_{+1}\\s_{-1}\end{pmatrix} = \begin{pmatrix}T_{11} & T_{12} \\ T_{21} & T_{22} \end{pmatrix} \begin{pmatrix}s_{-2}\\s_{+2}\end{pmatrix} 
\end{equation}
and is related to the $S$ matrix via the transformation:
\begin{equation}
    T = \frac{1}{S_{21}}\begin{pmatrix}1 & -S_{22} \\ S_{11} & -\det(S) \end{pmatrix}.\label{eq:S-to-T}
\end{equation}
In terms of the $T$ matrix, on the real-$\omega$ axis, realness is expressed as $T^*(i\omega)=T(-i\omega)$ and energy conservation as $\left|T_{11}\right|^2-\left|T_{21}\right|^2=\left|T_{22}\right|^2-\left|T_{12}\right|^2=1$, $T^*_{11}T_{12}=T^*_{21}T_{22}$, while reciprocity holds anywhere on the complex-$\omega$ plane and is written as $\det(T)=1$.

At a system complex pole $\omega_n$, there are non-zero outgoing fields ($s_{-1},s_{-2}\neq 0$) without an input ($s_{+1} = s_{+2} = 0$), so $T_{11}(\omega_n)=0$. Since then $D_{1n} \propto s_{-1}$ and $D_{2n} \propto s_{-2}$, the ports-coupling ratio of the mode is $\sigma_n = 1/T_{21}(\omega_n)$ and reciprocity further mandates $T_{12}(\omega_n)=-1/T_{21}(\omega_n)=-\sigma_n$.

The types of filters we are interested in satisfy $S_{22} = \gamma S_{11}$, namely $T_{12} = -\gamma T_{21}$. Therefore, for reciprocal such filters, we get $\sigma_n^2 = -T_{12}(\omega_n)/T_{21}(\omega_n) = \gamma$, as in Eq.~(\ref{r-equation}) of the main text. (Reminder that, if realness must hold, then $\gamma=\pm1$.)

Inversely, consider a unitary reciprocal system where all the modes satisfy $\sigma_n^2 = \gamma$. We write $S_{pq}(\omega) = A_{pq}(\omega)/P(\omega)$, where $P(\omega) = \prod_n(\omega-\omega_n)$ includes all the $2N$ poles $\omega_n$ and $A_{pq}(\omega)$ is a polynomial of degree at most $2N$, so Eq.~(\ref{eq:S-to-T}) implies that $T_{12}(\omega) = -A_{22}(\omega)/A_{21}(\omega)$ and $T_{21}(\omega) = A_{11}(\omega)/A_{21}(\omega)$. At a pole, we have $T_{12}(\omega_n) = -\sigma^2_{n}T_{21}(\omega_n)=-\gamma T_{21}(\omega_n)$, thus $A_{22}(\omega_n)-\gamma A_{11}(\omega_n)=0$ [because $A_{21}(\omega_n)\neq 0$]. Since the degree of $A_{22}-\gamma A_{11}$ is at most $2N$, we then have $A_{22}(\omega)-\gamma A_{11}(\omega) = \alpha P(\omega) \Leftrightarrow S_{22} = \gamma S_{11} + \alpha$, where $\alpha$ is a constant. Now, since $S$ is unitary, we have $|S_{22}|^2 = |\gamma S_{11} + \alpha|^2 = |S_{11}|^2 \Leftrightarrow |\alpha|^2+2\text{Re}[\alpha^*\gamma S_{11}(\omega)] = 0$ at \emph{all} real frequencies $\omega$, leading to $\alpha=0$ and thus $S_{22} = \gamma S_{11}$.

\section{NONSTANDARD TWO-PORT SPECTRA}
\label{app_nonstandard_spectra}

\begin{figure}[!h]
\includegraphics[width=\columnwidth,keepaspectratio]{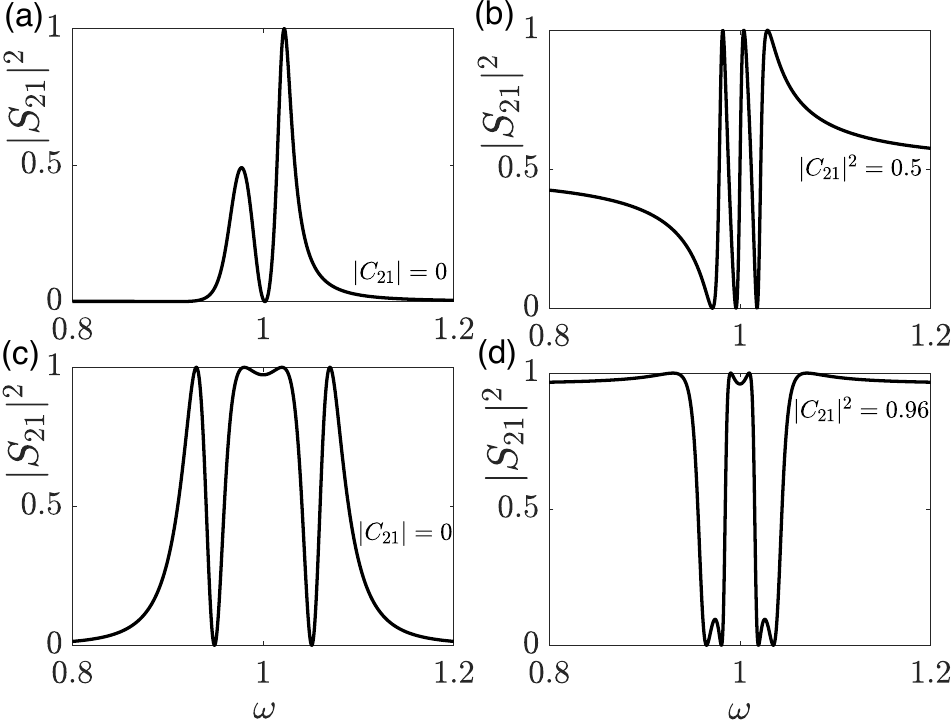}
\caption{Non-standard $S_{21}$ spectra of symmetric two-ports with $C=\begin{pmatrix}i r & t\protect\\t & i r \end{pmatrix}$ and resonances: (a) $\omega_n=(0.968-0.02i,0.99-0.02i,1.02-0.01i)$, $\sigma_n=(1,-1,-1)$, $t=0$, (b) $\omega_n=(0.98,1,1.02)-0.005i$, $\sigma_n=(1,1,1)$, $t^2=0.5$, (c) $\omega_n=(0.94,0.95,1.05,1.06)-0.02i$, $\sigma_n=(1,1,-1,-1)$, $t=0$, (d) $\omega_n=(0.96-0.01i,0.984-0.004i,1.016-0.004i,1.04-0.01i)$, $\sigma_n=(-1,1,-1,1)$, $t^2=0.96$.}
\label{Fig-nonstandard-spectra} 
\end{figure}

While standard filters are associated with alternating $\sigma_n$-signs and specific ``textbook'' poles, other choices can still give interesting spectra to design. QNMT allows for a quick computation of such spectra by simply plugging values for $\omega_n$, $\sigma_n$ and $C$. In \figref{Fig-C-background}, we showed possible \emph{non-standard} $S_{21}(\omega)$ spectra for $N=2$, and we provide some more examples for $N=3,4$ in \figref{Fig-nonstandard-spectra}. The general rules derived in \secref{sec_filter_types} still apply, so there is an odd/even number of zeros between modes of respectively same/opposite $\sigma_n$-sign and all spectra asymptote to $|C_{21}|$ outside the resonances' bandwidth.

\section{ANGLE DEPENDENCE OF THIRD-ORDER BANDPASS ELLIPTIC TRANSMISSION FILTERS}
\label{app_angle_dependence}

\begin{figure}
\includegraphics[width=\columnwidth,keepaspectratio]{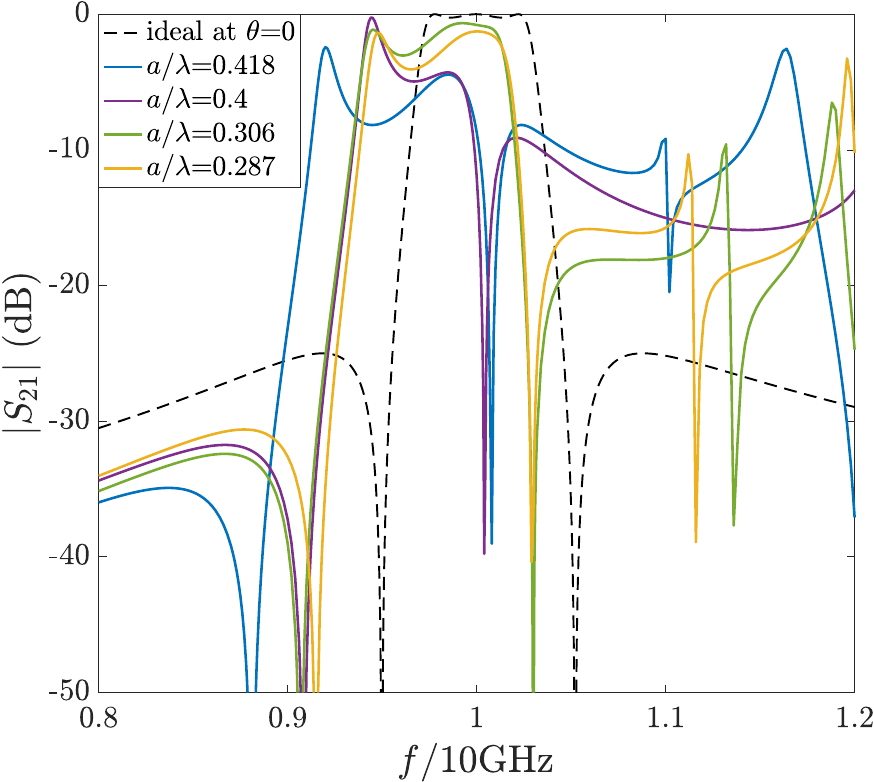}
\caption{TM-to-TM transmission at angle $\theta=15^\circ$ for the different optimized structures in Figs.~\ref{Fig-grid3}, \ref{Fig-optimizations} with ($\sim 6\%$ bandwidth) elliptic response at normal incidence. The angle response deviates more as the period increases.}
\label{Fig-angle-dependence} 
\end{figure}

We have shown four distinct physical designs for the same third-order elliptic transmission filter with $-0.25$ dB passband of $6\%$ bandwidth and with $-25$ dB stopbands: the red solid line in \figref{Fig-grid3} and the three red solid lines in \figref{Fig-optimizations}. Although their performance is by design very similar at normal incidence (matching the SF), they have different responses for non-zero off-axis angle $\theta$. In \figref{Fig-angle-dependence}, we show at $\theta=15^\circ$ along the $\Gamma\mathrm{X}_x$ line of the Brillouin zone the TM-to-TM transmission, which turns out to deviate \emph{more} from the designed ($\theta=0^\circ$) spectrum than the TE-to-TE transmission. (Note that, for incidence along $\Gamma\mathrm{X}_x$, TE and TM polarizations are still decoupled due to the $\hat{y}$-mirror symmetry.)

We see that the passband is shifted to lower frequencies, and higher-order modes get closer to the passband, reducing the stopband range. However, it can be observed that structures with smaller periodicity $a$, attained by using higher-$\epsilon$ dielectrics, tend to maintain better their performance. This matches common metasurface intuition, based on the rough principle that the parallel incidence wave vector $\omega\textrm{sin}\theta/c$ is a smaller fraction of the Brillouin zone edge at $\pi/a$. As a conclusion, if angle independence is important, a constrained optimization can be performed, with our filter-design criteria as constraints and some metric of this independence (e.g., flatness of resonant bands) as optimization objective.

\section{INDUCTIVE COUPLING BETWEEN CLOSELY SPACED APERTURES}
\label{app_mutual_inductance}

For two inductors $L_{1},L_{2}$ with mutual inductance $M$, the standard T-type coupling network with elements $L_{1}-M$, $L_{2}-M$ and $M$ is converted to the $\Pi$-type network, used in our circuit models, with element values $\left(L_{1}L_{2}-M^{2}\right)/\left(L_{2}-M\right)$, $\left(L_{1}L_{2}-M^{2}\right)/\left(L_{1}-M\right)$ and $\left(L_{1}L_{2}-M^{2}\right)/M$. The last element represents the longitudinal inductive coupling $L_{b}$ in our circuits, which becomes small for large mutual inductance $M<\sqrt{L_1L_2}$ (and vice versa).

When $L_{1},L_{2}$ are aperture-type, $M$ scales linearly with their on-axis distance $t$ as $M\approx M_{0}-\xi t$ for $\xi t\ll M_{0}$~\cite{conway2007inductance}, so $L_{b}\approx L_{1}L_{2}/M_{0}-M_{0}+\xi t$. When the two apertures are not too dissimilar, $M_{0}\approx\sqrt{L_{1}L_{2}}$, so we finally get $L_{b}\approx\xi t$ for $M_{0}\left(L_{1}L_{2}/M_{0}^{2}-1\right)\ll\xi t\ll M_{0}$. Moreover treating the dielectric layer as a very short transmission line, its equivalent circuit model is also just an inductor with $L'_{b}\approx\xi't$, where $\xi'$ the inductance per unit length. Combining the two sources of inductance we conclude the rough scaling $L_{b}\propto t$.

\begin{widetext}
\begin{center}
\begin{table}[!h]
\begin{tabular}{ | c c | c | c c c | c c c | c c c c | c c c c |} 
 \hline
 Type & BW$_{3\mathrm{dB}}$ & $a$ (mm) & $w_1/a$&$w_2/a$&$w_3/a$  & $l_1/a$&$l_2/a$&$l_3/a$ & $t_1/a$&$t_2/a$&$t_3/a$&$t_4/a$ & $\epsilon_1$&$\epsilon_2$&$\epsilon_3$&$\epsilon_4$ \\ \hline
 Butterworth & 5\% & 13.49 & 0.024 & 0.003 & 0.024  & 0.805 & 0.709 & 0.805 & --& 0.445 & 0.445 &-- & 1 & 2.41 & 2.41 & 1 \\ \hline
 Chebyshev & 5\% & 12.02 & 0.0099 & 0.0027 & 0.0099  & 0.7896 & 0.6639 & 0.7896 & --& 0.483 & 0.483 &-- & 1 & 3.45 & 3.45 & 1 \\ \hline
 Inv. Cheb. & 5.6\% & 9.83 & 0.221 & 0.050 & 0.055  & 0.772 & 0.645 & 0.944 & --& 0.278 & 0.096 & 0.018 & 1 & 6.72 & 3.64 & 2.73 \\ \hline
 Elliptic & 2.4\% & 10.00 & 0.28 & 0.034 & 0.009   & 0.501 & 0.535 & 0.849 & --& 0.451 & 0.026 & 0.020 & 1 & 8.21 & 3.95 & 3.00 \\ \hline
 Elliptic & 6\% & 9.175 & 0.222 & 0.068 & 0.022  & 0.692 & 0.607 & 0.908 & --& 0.343 & 0.071 & 0.021 & 1 & 8.58 & 4.42 & 3.19 \\ \hline
 Elliptic & 10.8\% & 10.45 & 0.207 & 0.012 & 0.071  & 0.806 & 0.710 & 0.992 & --& 0.187 & 0.066 & 0.014 & 1 & 5.72 & 2.61 & 3.41 \\ \hline \hline
 Elliptic & 6\% & 12.52 & 0.2395 &  0.110  &  0.0713  & 0.6117  &  0.5247  & 0.688 & -- &  0.379 & 0.090  & 0.0567 & 1 & 4.963 &  2.987 & 3.579 \\ \hline
Elliptic & 6\% & 12.00 & 0.0438  &  0.2704   & 0.0433  &  0.7291  &  0.7611 &   0.7009 & -- & 0.155 & 0.215 & 0.030 & 1 & 4.217 & 3.413 & 3.629 \\ \hline
Elliptic & 6\% & 8.619 & 0.233 &  0.163  & 0.0486  & 0.682  &  0.586  &  0.855 & -- & 0.336 &  0.117 &   0.02 & 1 & 10  & 5.898 &  3.2 \\ \hline
\end{tabular}
\caption{Structural parameters of metasurfaces implementing third-order bandpass filters of \figref{Fig-grid3} and of the QNMT-optimized structures in \figref{Fig-optimizations} (from top to bottom).}
\label{grid3_params}
\end{table}
\end{center}
\end{widetext}

\bibliography{biblio}{}

\pagebreak
\widetext
\begin{center}
\textbf{\large Supplemental Material}
\end{center}
\setcounter{equation}{0}
\setcounter{figure}{0}
\setcounter{table}{0}
\setcounter{page}{1}
\setcounter{section}{0}
\makeatletter
\renewcommand{\theequation}{S\arabic{equation}}
\renewcommand{\thefigure}{S\arabic{figure}}

\section*{QNMT parameters for structures}

Here, we provide all the QNMs computed via finite-element simulations. The computed QNM-to-CPM ratios are indicated by $\sigma$. The modes used to calculate the background $C$ matrix are marked in bold. $\Omega$ and $\Gamma$ in the tables are in units of $\omega_\mathrm{c}=2\pi\times10$~GHz.

We remind that good approximate solutions with complex $\gamma=e^{i\varphi}$ can be found, so we allow a common phase $\sqrt{\gamma}$ for the ratios $\sigma$ during optimization. Deviations of the final computed $\sigma/\sqrt{\gamma}$ from the ideal $\pm 1$ or $\pm i$ shown below lead to only small errors in the SF designs. For symmetric structures, all computed $\sigma$ are equal to $\pm 1$ anyway. (Note that, after the design optimization process is completed, we do not care to fine-tune $\sigma$ for QNMT modeling, since in Figs.~3-7 we only show the exact transmission spectra from direct frequency-domain simulations anyway.)\\

\textbullet\ \textbf{Second-order bandpass symmetric filters of Fig.~3}\\
Note that the zero is always on the side of the smallest $\Gamma$.

\begin{center}
\begin{tabular}{ | c | c | c || c | c | } 
 \hline
 Type & \multicolumn{2}{c||}{Zero on Left} & \multicolumn{2}{c|}{Zero on Right} \\ \hline
 $\Omega$ & 0.973 & 1.013 
 & 0.981 & 1.024 \\
 \hline
 $\Gamma$ & 0.0105 & 0.0234 
 & 0.0224 & 0.0110\\
 \hline
 $\sigma$ & -1 & 1 
 & 1 & -1 \\
 \hline
\end{tabular}
\end{center}

\bigskip{}

\textbullet\ \textbf{Third-order bandpass filters of Fig.~4}

\begin{center}
\begin{tabular}{ | c | c | c | c || c | c | c || c | c | c |} 
\hline
Type & \multicolumn{3}{c||}{Butterworth ($\varphi=0$)} & \multicolumn{3}{c||}{Chebyshev ($\varphi=0$)} & \multicolumn{3}{c|}{Inverse Chebyshev ($\varphi=-0.07\pi$)} \\ \hline
$\Omega$ & 0.9774 & 1.0010 & 1.0210 & 0.9788 & 1.0009 & 1.0223 & 0.9728 & 0.9975 & 1.0254 \\ \hline
$\Gamma$ & 0.0127 & 0.0251 & 0.0123 & 0.0079 & 0.0154 & 0.0075 & 0.0119 & 0.0341 & 0.0128 \\ \hline
$\sigma/\sqrt{\gamma}$ & 1 & -1 & 1 & 1 & -1 & 1 & -0.94+0.00i & 1.05+0.02i & -0.98+0.03i \\ \hline
\end{tabular}
\end{center}

\begin{center}
\begin{tabular}{ | c | c | c | c || c | c | c || c | c | c |} 
\hline
Type & \multicolumn{3}{c||}{Elliptic 2\% ($\varphi=-0.066\pi$)} & \multicolumn{3}{c||}{Elliptic 6\% ($\varphi=-0.059\pi$)} & \multicolumn{3}{c|}{Elliptic 10\% ($\varphi=-0.075\pi$)}  \\ \hline
$\Omega$ & 0.9885 & 0.9995 & 1.0101 & 0.9718 & 0.9987 & 1.0274 & 0.9516 & 0.9972 & 1.0504 \\ \hline
$\Gamma$ & 0.0032 & 0.0086 & 0.0028 & 0.0076 & 0.0218 & 0.0078 & 0.0124 & 0.0392 & 0.0145 \\ \hline
$\sigma/\sqrt{\gamma}$ & -1.04-0.14i &  1.16+0.03i &  -1.10+0.17i & -1.04+0.03i  & 0.91-0.01i & -1.01-0.04i & -1.04-0.15i & 1.19+0.00i & -1.03+0.15i \\ \hline
\end{tabular}
\end{center}

\bigskip{}

\textbullet\ \textbf{Third-order elliptic bandstop filter of Fig.~5(a)} ($\varphi=-0.08\pi$) 

\begin{center}
\begin{tabular}{ | c | c | c | c | } 
 \hline
 $\Omega$ & 0.944 & 0.999 & 1.061 \\
 \hline
  $\Gamma$ & 0.0155 & 0.0808 & 0.0159  \\
 \hline
 $\sigma/\sqrt{\gamma}$ & 1.05+0.04i & -1.03+0.00i  & 1.02-0.05i \\
 \hline
\end{tabular}
\end{center}

\bigskip{}

\textbullet\ \textbf{Second-order elliptic bandpass symmetric filter of Fig.~5(b)}\\
The modes marked in bold lead to $|C_{21}(\omega_\mathrm{c})|=-24.9$ dB and $\beta=C_{11}C_{21}^*/|C_{11}C_{21}|=-i$.

\begin{center}
\begin{tabular}{ | c | c | c | c | c | c | c | c | c | c | c | c | } 
 \hline
 $\Omega$ & \textbf{0.738} & \textbf{0.751} & 0.996 & 1.004 & \textbf{1.186} & \textbf{1.191} & \textbf{1.289} & \textbf{1.362} \\
 \hline
 $\Gamma$ & \textbf{0.297} & \textbf{0.301} & 0.0031 & 0.0031 & \textbf{0.0012} & \textbf{0.011} & \textbf{0.0084} & \textbf{0.0007} \\
 \hline
 $\sigma$ & \textbf{1} & \textbf{-1} & -1 & 1 & \textbf{-1} & \textbf{1} & \textbf{-1} & \textbf{1}\\
 \hline
\end{tabular}
\end{center}

\newpage

\textbullet\ \textbf{Second-order elliptic bandstop symmetric filter of Fig.~5(c)}\\
Note that the ``Fabry-Perot'' modes here have such a large $\Gamma$ that higher-order such modes will still have an effect at frequencies around $\omega\sim \omega_\mathrm{c}$. However, they lie deep inside the diffraction zone, so they are hard to identify from spurious modes. Therefore, as explained in the main text, we calculated more accurately $C(\omega_\mathrm{c})=\bar{S}^{-1}(\omega_\mathrm{c})S(\omega_\mathrm{c})$, with $\bar{S}(\omega_\mathrm{c})$ obtained from QNMT on the two high-$Q$ (non-bold) modes and $S(\omega_\mathrm{c})$ from a direct computation. The optimized result is then indeed $|C_{21}(\omega_\mathrm{c})|=-0.25$ dB and $\beta=i$.

\begin{center}
\begin{tabular}{ | c | c | c | c | c | c | c | c | c | c | c | c | } 
 \hline
 $\Omega$ & \textbf{0} & 0.956 & 1.044 & \textbf{1.111} & \textbf{1.462} & \textbf{1.483}& \textbf{1.896}\\
 \hline
  $\Gamma$ & \textbf{0.853} & 0.0313 & 0.0343 & \textbf{0.995} & \textbf{0.0146} & \textbf{0.0011} & \textbf{0.956} \\
 \hline
 $\sigma$ & \textbf{1} & -1 & 1 & \textbf{-1} & \textbf{1} & \textbf{-1} & \textbf{1} \\
 \hline
\end{tabular}
\end{center}

\bigskip{}

\textbullet\ \textbf{Third-order Chebyshev bandpass symmetric filter of Fig.~6}

\begin{center}
\begin{tabular}{ | c | c | c | c | } 
 \hline
 $\Omega$ & 0.9727 & 0.9999 & 1.0272 \\
 \hline
  $\Gamma$ & 0.00948 & 0.0192 & 0.00970  \\
 \hline
 $\sigma$ & 1 & -1  & 1 \\
 \hline
\end{tabular}
\end{center}

\bigskip{}

\textbullet\ \textbf{Second-order elliptic bandpass symmetric filter of Fig.~7}\\
For this diffractive structure, there are several other high- and low-$Q$ QNMs nearby (not listed in the table), plus branch points of the -1 and higher orders. Therefore, instead of attempting a direct QNMT computation of $C$, we resort again to $C=\bar{S}^{-1}S$. Due to approximation errors, $C$ exhibits small oscillations very close to $\omega_\mathrm{c}$, so, instead, we calculate it at frequencies $(0.959,0.977,1.023,1.041)\omega_\mathrm{c}$ and interpolate to find $|C_{21}(\omega_\mathrm{c})|=-25.1$ dB. During optimization, we used this interpolation to also minimize $d|C_{21}(\omega_\mathrm{c})|/d\omega$ and  $d^2|C_{21}(\omega_\mathrm{c})|/d\omega^2$ to achieve a very slow variation of $C$.

\begin{center}
\begin{tabular}{ | c | c | c | c | c | c | c | c | c | c | c | c | } 
 \hline
 $\Omega$ & 0.9941 & 1.0059\\
 \hline
 $\Gamma$ & 0.00437 & 0.00442\\
 \hline
 $\sigma$ & -1 & 1\\
 \hline
\end{tabular}
\end{center}

\end{document}